\documentclass[twocolumn,final,natbib]{svjour3}
\usepackage{fixltx2e}          % fixes two-column figure numbering

\usepackage{graphicx}          % Almost always needed
\usepackage{mathptmx}      % use Times fonts if available on your TeX system

\usepackage[colorlinks,citecolor=blue,linkcolor=blue,urlcolor=blue,pdftex]{hyperref}
\usepackage{journalabbrevs}
\usepackage{doi}  % makes DOIs in bibliography into links (plainnat only)

\graphicspath{{./figs/}}
\newcommand{\degr}{^\circ}
\newcommand{\degW}{^\circ\,\mathrm{W}}
\newcommand{\degE}{^\circ\,\mathrm{E}}
\newcommand{\degN}{^\circ\,\mathrm{N}}
\newcommand{\hPaunit}{\,\mathrm{hPa}}
\newcommand{\munit}{\, \mathrm{m}}
\newcommand{\windunit}{\, \mathrm{m\,s}^{-1}}
\newcommand{\trendunit}{\, \mathrm{m}\,\mathrm{s}^{-1}\,\mathrm{decade}^{-1}}
\newcommand{\Uerai}{U^\mathrm{ERAI}}
\newcommand{\Utwcr}{U^\mathrm{20CR}}
\newcommand{\Utwcrbc}{U^\mathrm{20CRc}}

\newcommand{\figw}{84mm} % figure width for 2-column formatting
\newcommand{\figbig}{174mm} % figure width for 2-column formatting
\journalname{Theoretical and Applied Climatology}
\begin{document}
%=============================================================================
\title{Using the Twentieth Century Reanalysis to assess climate variability  for the European wind industry}
\titlerunning{Using 20CR for European winds}    % if too long for running head
\author{Philip E. Bett \and
        Hazel E. Thornton \and
        Robin T. Clark}
\authorrunning{Bett, Thornton, Clark} % if too long for running head

\institute{P. E. Bett \and H. E. Thornton \and R. T. Clark
  \at Met Office Hadley Centre, FitzRoy Road, Exeter, EX1~3PB \\
              \email{philip.bett@metoffice.gov.uk}           %  \\
}

\date{Received: date / Accepted: date\\
\noindent\copyright\ Crown Copyright 2015, the Met Office. Author's corrected version.}
% The correct dates will be entered by the editor.

\maketitle
%=============================================================================
\begin{abstract}

We characterise the long-term variability of European near-surface wind speeds using  142  years of data from the Twentieth Century Reanalysis (20CR), and consider the potential of such long-baseline climate data sets for wind energy applications. 
The low resolution of the 20CR would severely restrict its use on its own for wind farm site-screening.  We therefore perform a simple statistical calibration  to link it to the higher-resolution ERA-Interim data set (ERAI), such that the adjusted 20CR data has the same wind speed distribution at each location as ERAI during their common period.  
Using this corrected 20CR data set, wind speeds and variability are characterised in terms of the long-term mean, standard deviation, and corresponding trends. Many regions of interest show extremely weak trends on century timescales, but contain large multidecadal variability. 
Since reanalyses such as ERAI are often used to provide the background climatology for wind farm site assessments, but contain only a few decades of data, our results can be used as a way of incorporating  decadal-scale wind climate variability into such studies, allowing investment risks for wind farms to be reduced.
%\keywords{windFirst keyword \and Second keyword \and More}
% \PACS{PACS code1 \and PACS code2 \and more}
% \subclass{MSC code1 \and MSC code2 \and more}
\end{abstract}

%=============================================================================

\section{Introduction}\label{s:intro}
Wind is a highly variable phenomenon over all time scales, from gusts lasting seconds, to long-period variations spanning decades \citep[e.g.][]{Watson2014Quantifying}.  Harnessing the wind resource for electricity production is a rapidly-developing field, with many challenges for engineering, energy systems design, national-scale energy policy, and meteorological forecast systems \citep[e.g.][]{srrench7wind}.  Short term wind variability is critically important to the  day-to-day management of a wind farm, and efficient running depends on having high quality wind speed forecasts \citep[e.g.][]{Foley2012Current, Jung2014Current}.  However, the impact of long term, decadal-scale variations in the wind climate is less well understood.

This is partly due to a historical lack of data.  Typically, when a site is considered for wind farm development, developers are often restricted to using statistical techniques to relate observational records from nearby stations to the site in question. Homogeneous data from any single station will usually only span a few years to a decade, but can be supplemented by data from a dedicated meteorological mast positioned on-site for a limited period of time such as 1--3 years \citep{Petersen2012Wind, elforsk2013, Carta2013MCPReview}.  In the absence of long term data sets of wind speed itself, studies of long term wind variability typically use pressure-based metrics as proxies for the wind \citep[e.g.][]{Palutikof1992ukwind}, often combined with complex statistical procedures to relate to the wind speed at a site \citep[e.g.][]{KirchnerBossi2013, KirchnerBossi2014Longterm}.  Around Europe,  indices based on the North Atlantic Oscillation (NAO) have often been  used \citep[e.g.][]{Boccard2009Capacity}.   Standard  NAO indices  correlate well with winter wind speeds in northern/western parts of Europe.  However, this is not true more generally, such as at other times of the year or in other locations \citep{Hurrell2003Overview}, and alternative indices must be used in these cases \citep[e.g.][]{Folland2009Summer}.  Regardless of definition, the NAO does not capture the full variability seen in wind speeds.   Thus, there is scope for improvement over all these techniques.

Within the past decade, reanalysis data products have been able to extend such site assessment studies, allowing a description of a reasonable climatological period of around 30 years.  The two main global reanalysis data sets used for this are the \emph{ECMWF}\footnote{European Centre for Medium-range Weather Forecasting} \emph{Re-Analysis Interim} product (ERA-Interim, hereafter ERAI; \citealt{Dee2011}), and NASA's \emph{Modern Era Retrospective-analysis for Research and Applications} (MERRA, \citealt{Rienecker2011MERRA}), which both cover the `satellite era' (1979 onwards).  Such data sets are necessarily produced at relatively low spatial resolution (e.g. grid sizes  $\sim 0.7\degr$), and cannot, on their own, be used to determine the likely wind speeds at a site.  In combination with other techniques however, from simple rescaling,  detailed statistical modelling or even full dynamical downscaling, reanalysis data can be a key source for obtaining a representative wind climatology for a specific site \citep{Kiss2009Comparison, Petersen2012Wind, Kubik2013Exploring,Badger2014Wind}. 

Most recently, attempts at producing century-scale reanalyses have yielded results: the NOAA\footnote{National Oceanic and Atmosphere Administration} \emph{Twentieth Century Reanalysis} \citep[hereafter 20CR,][]{Compo2011} and ECMWF's ERA-20C \citep{Poli2013era20c, Dee2013Toward} data sets provide ensemble realisations of the atmospheric state spanning over 100 years.  However, as they are at even lower resolution (e.g. $1$--$2\degr$), and their early data is subject to substantial uncertainty, care must be taken when considering how to interpret their results in the context of wind farms.

Concerns within the wind industry about the possible impacts of future climate change, along with greater availability of larger data sets,  have  motivated various studies resulting in a greater awareness of the risks of climate variability (whether anthropogenic or natural).  In fact, unlike the situation for temperature, there is little evidence of any long-term trend in globally-averaged wind speeds -- see e.g. the Fourth and Fifth Assessment Reports (AR4/AR5 respectively) of the IPCC's\footnote{Intergovernmental Panel on Climate Change} Working Group I, \cite{ipcc2007wg1ch3} and \cite{ipcc2013wg1ch2}.  The low confidence in such assessments is due in part to difficulties with the historical observational record, coupled with the highly-variable nature of winds in both space and time.  For example, various data sets have suggested a positive trend in wind speeds over the oceans, with significant regional variability \citep{Tokinaga2011Wave, YoungCommentReply2011, Young2011Global, WentzComment2011, Young2012Investigation}.  Over land however, the situation is different: an apparent reduction in surface wind speeds (nicknamed ``global stilling'') has been seen in recent decades in some data sets \citep{McVicar2012, sotc2012globalwinds}, with studies suggesting that it could be due in part to anthropogenic factors, such as changes in land-use  increasing the surface roughness \citep{Vautard2010, Wever2012}, or aerosol emissions locally changing the thermal structure of the atmosphere \citep{Bichet2012}.  It is important to note that stilling is not seen in reanalysis data, which use climatological aerosol levels and do not include land-use change.  Over both the land and oceans, opposing trends in different regions and times of year will act to reduce any globally-averaged trend signal.   While further and better data is required to settle questions on the true scale, causes and interrelationships of changes in wind speeds over oceans and land, it is important to note that these observed trends are always much smaller than  interannual variability.

Given the uncertainties in trends in the historical wind climate, it is not surprising that  projections of future wind climates should also be treated with caution.  The review of \cite{Pryor2010Climate} concluded that wind speeds over Europe would continue to be dominated by natural variability, although by the end of the century some differences could have emerged -- although even the sign of the change was uncertain.  The IPCC's Special Report on Renewable Energy Sources and Climate Change Mitigation (SRREN) came to a similar conclusion \citep{srrench7wind}, and the IPCC's AR4 \citep{ipcc2007wg1ch10, ipcc2007wg1ch11} and AR5 \citep{ipcc2013wg1ch12, ipcc2013wg1ch14} noted that there is low confidence in any projected changes.  Consequently, \cite{Pryor2010sres} and \cite{Dobrynin2012Evolution} found that the choice of emission scenario or concentration pathway has relatively little impact overall on the resulting wind climate.
It is important to note that simulations of the historical climate over the 20th Century (from both atmosphere-only and ocean-coupled models) do not reproduce the observed variability in atmospheric circulation \citep{scaife2005, Scaife2009CLIVAR}, so the uncertainties in these climate projections do not preclude large multi-decadal variations in the future.

Overall, the effect of climate change on the annually-averaged wind resource is thought to be small, although the increased seasonality seen in some studies by 2100 could have a challenging impact on wind-dominated electricity networks  \citep{Hueging2012Regional, Cradden2012}.

Thus, when seeking to improve assessments of future wind speeds over the lifetime of a turbine, there is more to be gained from an increased understanding of historical long-term wind variability than through climate change model runs.   Given this context, we show in this paper how the new class of century-scale reanalyses can be linked to the more widely-used satellite-era reanalyses, thus allowing for information on the long-term decadal-scale variability in wind speeds to be propagated through the model chain when performing a wind site assessment.   In Section~\ref{s:datasources} we describe the two main data sets we use, including their limitations.  We compare them in detail in Section~\ref{s:reanallinking}, and describe our procedure for relating the two.  Section~\ref{s:results} shows results for the wind speed distribution over Europe, including long-term averages, variabilities, and changes in the shape of the distribution over time for selected regions.  We discuss our conclusions in Section~\ref{s:conclusions}.

%===========================================================

%===========================================================

\section{Data sources}\label{s:datasources}

Reanalyses represent the most convenient data sets for assessing the long-term historical wind climate, in the sense that they aim to provide an optimal combination of observations and numerical model: the data provided in a reanalysis aims to give the best estimate of the ``true'' situation at any given point, as well as being homogeneous in time (e.g. free of systematic shifts),  and complete in both space and time.  However, in reality, biases and uncertainties inherent in both raw observations (due to location, frequency, instrumentation, etc.) and models (due to resolution, parametrisation schemes, etc.) mean that such data sets must be used with caution.

This study primarily uses data from the \textit{Twentieth Century Reanalysis} project (20CR), in conjunction with  wind speeds from the ERA-Interim  data set (ERAI) for validation and calibration of the 20CR data.  We describe key aspects of these data sets in the following sections.

\subsection{Data from the 20CR ensemble system}\label{s:twcr}
A full description of the ensemble reanalysis system used in the 20CR project is given in \cite{Compo2011}.  Here, we describe some key features that have  important impacts on our analysis methods and results.  

The 20CR assimilates sea-level pressure and surface pressure observations alone (from the International Surface Pressure Databank, incorporating the ACRE\footnote{Atmospheric Circulation Reconstructions  over the Earth, \url{http://www.met-acre.org/}} project, \citealt{ACRE2011}), using observational fields of sea-surface temperature and sea-ice concentration  (HadISST1.1, \citealt{Rayner2003}) as boundary conditions.  It uses the April 2008 experimental version of the NCEP\footnote{National Centres for Environmental Prediction} Global Forecast System (GFS), a coupled atmosphere--land model produced by the NOAA NCEP Environmental Modelling Centre.

The 20CR data assimilation system is based on an Ensemble Kalman Filter.  The data are produced in a series of 5-year\footnote{Streams 16 \& 17 actually last 6 and 4 years respectively (see Table III in \citealt{Compo2011}). For simplicity, we assume 5-year streams throughout.} `streams' (independent runs, to simplify parallelisation), with 56 members in each stream.
A consequence of this system is that ensemble members only remain temporally continuous for the 5-year duration of each stream.  This has implications for how variability is assessed over long time periods; we discuss this in more detail in Section~\ref{s:mapstrends}.

As highlighted in \cite{Compo2011}, when considering variability it is important to use the ensemble members directly, rather than using the daily ensemble-mean time series alone.  The increased uncertainty in the early period of the data leads to greater disagreement between the ensemble members, such that a time series of their mean will have much less variability than the members individually.  This would lead to a spurious strong reduction in variability appearing at earlier times in the ensemble mean.

We use the updated release of the \emph{20CRv2} data (hereafter simply 20CR), spanning 142 years from 1st Jan 1871 to 31st Dec 2012.  While it was produced  on a T62 spectral grid with 28 vertical levels, we use the output data provided on a regular latitude--longitude grid with cell size $2\degr$, at the the near-surface pressure level at $\sigma := P/P_\mathrm{surface} = 0.995$ (around $40\munit$ height).  The $\sigma=0.995$ level is a reasonable choice for turbines whose rotor hubs are expected to be some tens of metres above the surface; typical hub heights are between $40$ \& $100\munit$, but vary greatly \citep{srrench7wind}; we do not expect our conclusions to be qualitatively affected by the precise height above ground.    More details on our choice of levels can be found in Appendix~\ref{s:levels}.   We use daily-mean wind speeds $U$, which we calculate by averaging the wind speed magnitudes from the 6-hourly $u$ (zonal, i.e. westerly) and $v$ (meridional, i.e. southerly) component fields.  We are not considering sub-daily variability, as this is likely to be poorly represented with only four timesteps per day, in addition to the low horizontal resolution. Using daily means significantly reduces the amount of data that we need to analyse.   However, calculating daily means using only four snapshots is likely to lead to some underestimation, as the wind distributions we are sampling tend to be positively skewed.  Using daily means also has an impact on the form of the resulting wind speed distribution, and on Weibull fits in particular; we discuss this in the Supplementary Information.

Some recent studies have highlighted potential problems with the 20CR data set.  \cite{Ferguson2012,Ferguson2014Evaluation} have performed a detailed analysis of change points in the 20CR data, finding that, while these are in fact common in the data set overall, there are many areas, especially in the northern hemisphere, where the 20CR remains largely homogeneous for many decades.  Their results emphasize that users of the 20CR data must be aware of possible -- indeed, \emph{probable} -- inhomogeneities in the data, and the potential impact this could have on their analyses. 
\cite{Stickler2011} found very significant differences between 20CR winds and pilot balloon measurements in the West African Monsoon region over 1940--1957, and \cite{elforsk2013}, using the 20CR to study interannual wind variability over Scandinavia, had to discard 20CR data prior to 1920 due to suspicious behaviour in some grid cells.
Finally,  there has been some debate on the consistency of long-term trends in storminess and extreme winds found in 20CR compared to observations (\citealt{Donat2011Reanalysis}, \citealt{Bronnimann2012Extreme}, \citealt{Wang2013Trends,Wang2014Storminess}, and \citealt{Krueger2014Comment, Krueger2013Inconsistencies}).    These studies serve to emphasize the importance of being extremely careful  with methodology when comparing reanalysis data with observations, and when identifying trends.

\subsection{Data from ERA-Interim}\label{s:era}
The second source of data we use is the $60\munit$ wind speed fields from the ERAI data set \citep{Dee2011}.  This uses the ECMWF Integrated Forecasting System model (IFS), and assimilates observational data of many types, mostly coming from satellites.  The atmospheric fields of ERA-Interim were calculated on a T255 spectral grid, with surface fields calculated on a reduced Gaussian grid.  We use the 6-hourly wind speed data available on the regular latitude--longitude grid of cell-size $0.75\degr$, and calculate daily-mean wind speeds as for the 20CR.  A comparison of ERAI data at $60\munit$ and $10\munit$ with the 20CR levels can be found in Appendix~\ref{s:levels}. The reanalysis starts in 1979 and continues to the present; we use the data up to the end of 2013.  Further details are available in \cite{Dee2011} and references therein, and the ERA-Interim Archive report \citep{Berrisford2011ERAInterim}.

\cite{Stopa2014Intercomparison} compared ERAI wind speeds with those measured from buoys and satellite data, finding that the reanalysis  performs very well in terms of homogeneity, but with a small negative bias and reduced variability compared to the observations.  \cite{Szczypta2011Verification}  found that ERA-Interim tended to overestimate wind speeds over most of France, but underestimated it in mountainous areas, compared to the SAFRAN high resolution (8\,km) reanalysis data set -- although the authors note that the SAFRAN wind speed data is known to be biased low.  

As already discussed, it is known that reanalysis data sets including ERA-Interim do not exhibit the observed large-scale trends in wind speeds (see e.g. \citealt{sotc2012globalwinds, sotc2012oceanwinds} and references therein), and the relatively low resolution of ERAI (and similar data sets) prevents it from being used directly as a proxy for observations at the scale of a wind farm \citep{Kiss2009Comparison, Kubik2013Exploring}.  We will instead be using  ERAI as an example of the kind of data currently used for  providing a climatological basis for wind farm site assessments, the first link in the `model chain' of dynamical and statistical downscaling for such studies: reanalyses are connected to mesoscale dynamical models, then in turn to microscale models and computational fluid dynamics (CFD) at the scale of a wind farm itself \citep{Petersen2012Wind}.

%===========================================================

%===========================================================

\section{Linking the reanalyses}\label{s:reanallinking}
While the strength of the 20CR is its characterisation of real-world variability on long time scales, the ERA-Interim data set provides wind speeds that are at much higher spatial resolution, and are more tightly-constrained by observations.  ERA-Interim is therefore much better suited for developing a climatology of wind speeds over small (sub-national) regions, or, in conjunction with additional dynamical or statistical downscaling techniques, at a point location.  However, as it only spans $\sim 30$ years it cannot give a good indication of climate variability on multi-decadal timescales.  In this section we describe how we calibrate the 20CR wind speed data to produce a data set that has the same distribution of wind speeds in time as ERA-Interim (over their overlapping period), but with the long-term variability of 20CR.

\subsection{Comparison of the reanalyses}\label{s:comparison}
We focus our study on Europe, and consider several small sub-regions for more detailed examination.  To aid comparison, we regrid the ERAI data by area-averaging onto the 20CR's native $2\degr$ grid.

The 20CR and ERAI data do not exhibit the same climatology in wind speeds over their period of intersection (1979--2012, 34  years).  This is due to a number of factors.  These include the structural differences (NWP model, data assimilation and reanalysis procedure); spatial resolution and the amount of orographic complexity resolved;  the amount and type of observational data assimilated; and the mismatch between vertical levels available for comparison.  
 
In this section we denote ensemble-mean daily-mean wind speeds from 20CR (at its  $\sigma=0.995$ vertical level) and from ERAI (at its $60\;\mathrm{m}$ model level on the 20CR grid), by $\Utwcr$ and $\Uerai$ respectively.  As we are focusing on the later period of the 20CR data set for our calibration procedure, the ensemble spread is small, so it is acceptable to use the ensemble mean series in this case (this is not generally true for all time periods, or regions of the globe with fewer observations; see \citealt{Compo2011}).  We consider the `bias' between the 20CR and ERAI data in terms of the simple difference in wind speeds, 
\begin{equation}
\beta := \Utwcr - \Uerai,
\end{equation}
and the day-to-day relative difference compared to ERAI,
\begin{equation}
\beta_\mathrm{rel} := \left(\Utwcr-\Uerai\right) / \Uerai.
\end{equation}

\begin{figure} \centering
  \includegraphics[width=\figw]{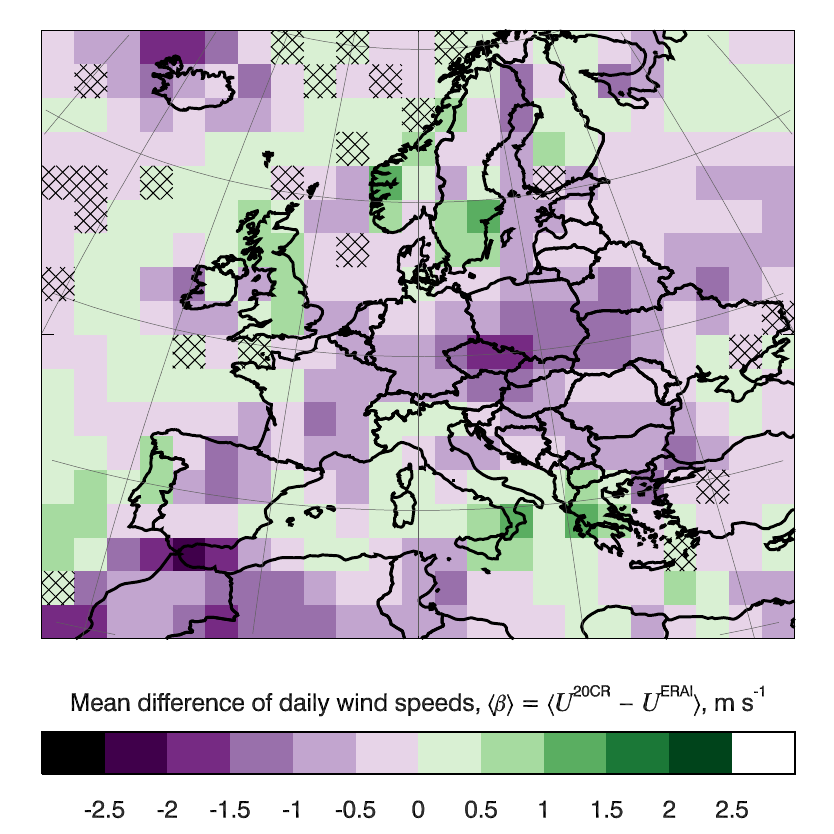} %{calib_meandiffmap}
  \includegraphics[width=\figw]{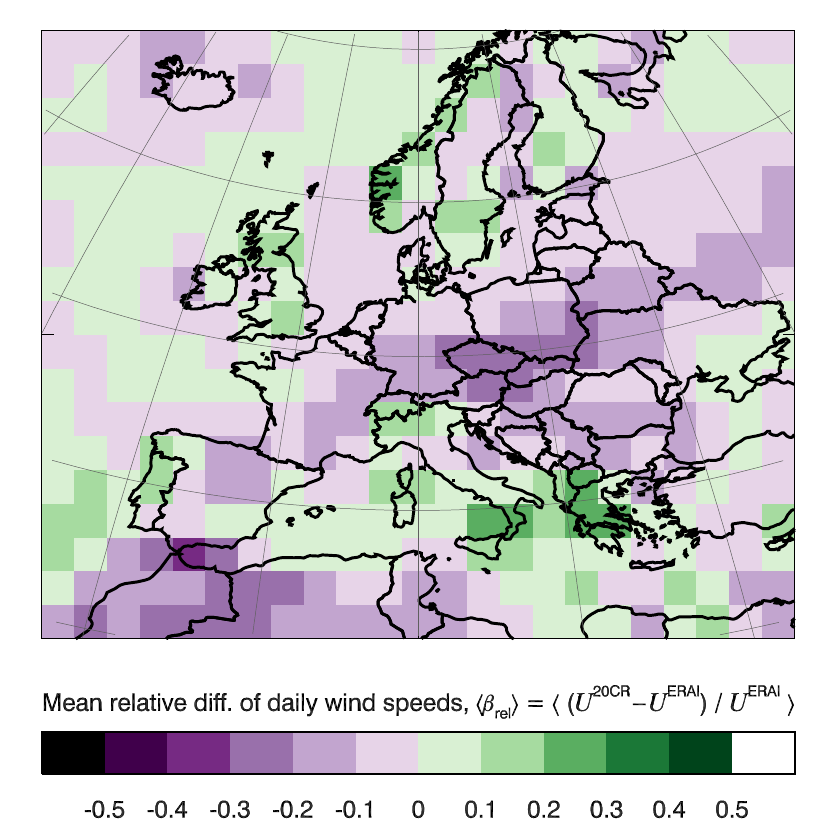} %{calib_meanfdevmap}
  \caption{Maps of the difference between wind speeds from 20CR and ERA-Interim; details as given in the panels.  Crosshatched areas in the top panel are not significantly different from zero at the $1\%$ level, according to a $t$-test.}
 \label{f:meanbiasmaps}
\end{figure}

In Fig.~\ref{f:meanbiasmaps}, we show\footnote{Throughout this paper we present maps on the 20CR's $2\degr$ grid in a Lambert Azimuthal Equal-Area projection centred on  ($10\degE$, $52\degN$), following e.g. \cite{annoni2003}, code \href{http://epsg-registry.org/}{EPSG::3035}. Calculations are performed on the regular lat.--lon. grid.} the  34-year mean bias $\langle \beta \rangle$  and the mean of the day-to-day relative bias $\langle\beta_\mathrm{rel}\rangle$.  The bias maps are all rather noisy, but over most of the land surface the bias is negative (i.e. $\Utwcr < \Uerai$), with differences of up to $\sim 20\%$ of the ERAI wind speeds in many areas.  There are some notable exceptions to this however, with positive biases (i.e. $\Utwcr > \Uerai$): for example over Britain, wind speeds are up to $20\%$ higher in the 20CR data.  Some areas have particularly strong negative bias, such as  around the Czech Republic.  The Strait of Gibraltar is particularly affected by the low spatial resolution, resulting in the lowest 20CR wind speeds compared to ERAI.   We have used a $t$-test to assess whether the data is consistent with  $\langle \beta \rangle = 0$ (i.e. no bias) at the $1\%$ level.  When it is not consistent with zero,  we say there is a significant bias; this is the case for most areas according to this test.

% They can be bigger as a column, but tidier as a row?
%\newcommand{\biasmapw}{78mm}           % column
%\newcommand{\biasmapw}{0.33\textwidth}  % row

\begin{figure}
\includegraphics[width=\figw]{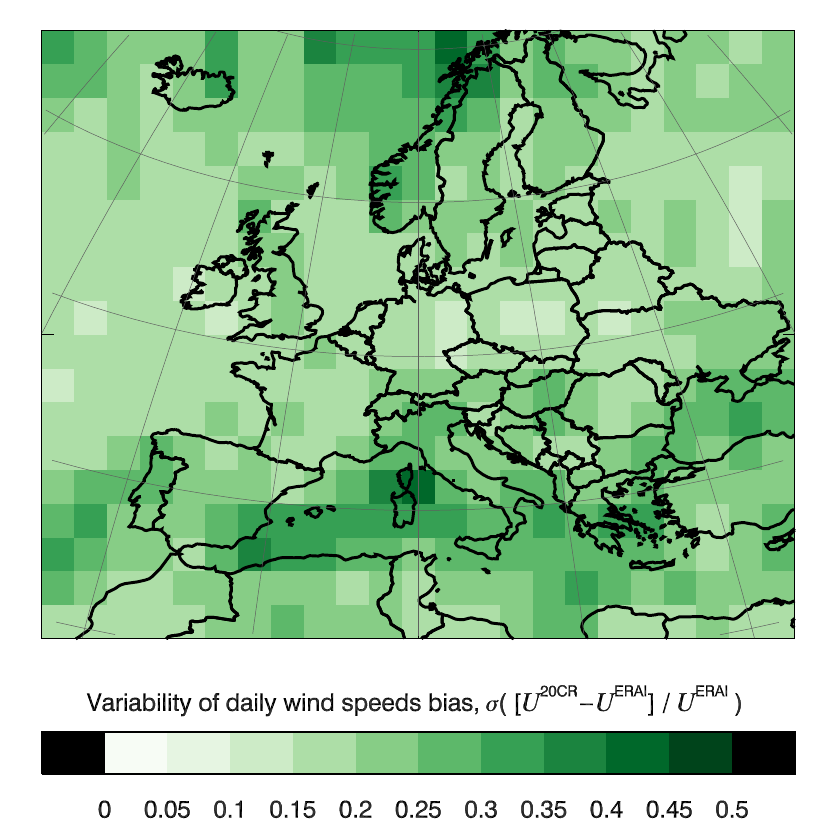} %{calib_stdvfdevmap}
\caption{Map of the variability of the daily relative `bias' $\beta_\mathrm{rel}$ between 20CR and ERAI wind speeds, in terms of its standard deviation.}
\label{f:biasvarmaps}
\end{figure}

In addition to the spatial variability, it is important to bear in mind that the difference between 20CR and ERAI does not have to be constant in time.   Fig.~\ref{f:biasvarmaps} shows the day-to day variability of  $\beta_\mathrm{rel}$ in terms of its  standard deviation $\sigma$.
There is a suggestion in the data of increased $\beta_\mathrm{rel}$ around coastal regions, such as in large parts of the Mediterranean, as well as Norway and Britain.
The relative-bias variability is generally around 15--30\%, which is a similar magnitude to the mean relative bias  $\langle \beta_\mathrm{rel} \rangle$ shown in Fig.~\ref{f:meanbiasmaps}. 

\begin{figure} 
\includegraphics[width=\figw]{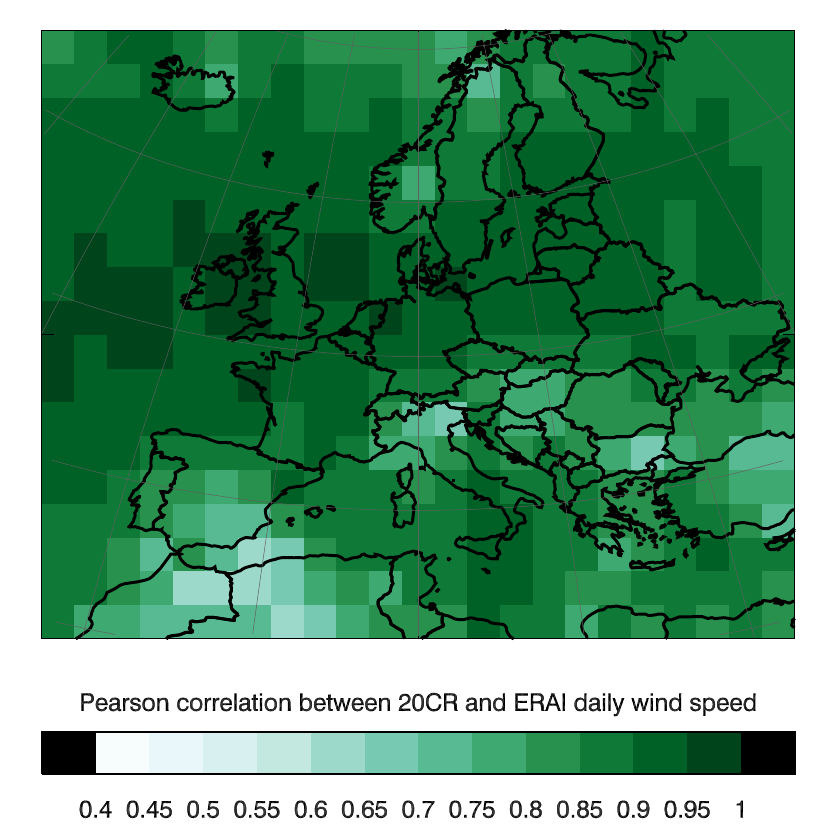} %{calib_correlmap}
\caption{Map of the Pearson correlation between daily-mean wind speeds in 20CR and ERA-Interim.  }
\label{f:correlmaps}
\end{figure}

Finally, we show the correlation between the daily wind speeds of the 20CR and ERAI in Fig.~\ref{f:correlmaps}.  The data are well-correlated in most places, but the correlation is particularly strong ($\geq 0.9$) in the Atlantic and northern Europe, including the British Isles.   

\subsection{Procedure for calibration}\label{s:biascorrectionmethod}
The goal of our calibration procedure is to generate a wind speed data set that retains the  fluctuation patterns of the 20CR data over time, and between ensemble members, but whose probability density functions (PDFs) of the ensemble-mean wind speed in each grid cell match those of the ERA-Interim data during their overlapping time period.  In particular, the PDFs do not have to match over other periods (e.g. if comparing the distribution over 142  years from 20CR to the 35 years from ERAI), the time series do not have to match in detail (although we have shown that they do tend to be well-correlated), and individual ensemble members do not need to match ERAI data -- thus retaining the 20CR's important measure of uncertainty.  We illustrate our procedure for the case of a particular grid cell in Fig.~\ref{f:calibmethod}.

Our method proceeds in two stages, and is performed on each grid cell independently.  Firstly a transfer matrix is obtained as the conditional probability density of ERAI wind speeds, given bins of 20CR ensemble mean, daily mean wind speeds for the overlapping period:
\begin{equation}
  \label{e:probmatrix}
  \mathcal{P}_{ij} := P(\Uerai_i|\Utwcr_j),
\end{equation}
where $i$ and $j$ index bins in wind speed for the data sets indicated.  We use bins of $0.5\windunit$ covering the range $0$--$40\windunit$.  This transfer matrix is applied to the full 142-year 20CR PDF,  to obtain a calibrated PDF of 20CR wind speeds spanning 1871--2012:
\begin{equation}
  \label{e:matrixmul}
  P(\Utwcrbc_i) = \sum_j \mathcal{P}_{ij} P(\Utwcr_j).
\end{equation}

Secondly,  calibrated daily time series of wind speeds, $\Utwcrbc(t)$,  from all ensemble members, are obtained by quantile matching \citep[e.g.][]{Panofsky1968Some}: 
the cumulative distribution function (CDF) of the calibrated 20CR ensemble mean wind speeds is interpolated at the quantiles of each ensemble member's wind speed (see bottom-left panel in Fig.~\ref{f:calibmethod}). Using the individual ensemble members in this step rather than the ensemble mean allows the ensemble spread to be transferred to the calibrated climatology.

In some cases, there can be wind speeds present in the 142-year 20CR data that were greater than any in the 34-year period common with ERAI.  This means that such wind speeds have no corresponding frequency in ERAI that we can calibrate to:  the CDF corresponding to $P(\Utwcrbc_i)$ reaches its maximum\footnote{Note that constructing the CDF in finite bins in wind speed, using a finite number of days, and relating the ensemble member time series to the ensemble mean distributions, means that sometimes the calibrated CDF does not quite reach unity.} below that wind speed, so quantile matching by interpolating the CDF will fail.  In this case, we instead perform a linear regression on the relationship between original and corrected winds up to this point (i.e.  $\Utwcrbc(\Utwcr) = a \Utwcr + b$). We then extrapolate this model to obtain corrected wind speeds for the final few high wind days. 

We demonstrate our procedure for the case of a grid cell in north-western Germany in Fig.~\ref{f:calibmethod}.  This shows the different PDFs in question, the transfer matrix, and the quantile matching.  It is clear that in this case the ERAI wind climate largely represents  a shift to higher wind speeds compared to 20CR (i.e. the 20CR winds are low compared to ERAI), and the calibrated 20CR reproduces this well.  The PDFs of both the ERAI and 20CR wind speeds appear somewhat truncated at lower wind speeds, rather than reducing smoothly towards $U=0\windunit$.  This is due to the daily averaging of the 6-hourly wind speeds, and has implications when attempting to fit Weibull functions to the wind speed distribution; we  discuss this issue in detail in the Supplementary Information.

\newcommand{\calibfourw}{0.49\columnwidth} %{0.35\textwidth}
\begin{figure} 
\centering
\includegraphics[width=\calibfourw]{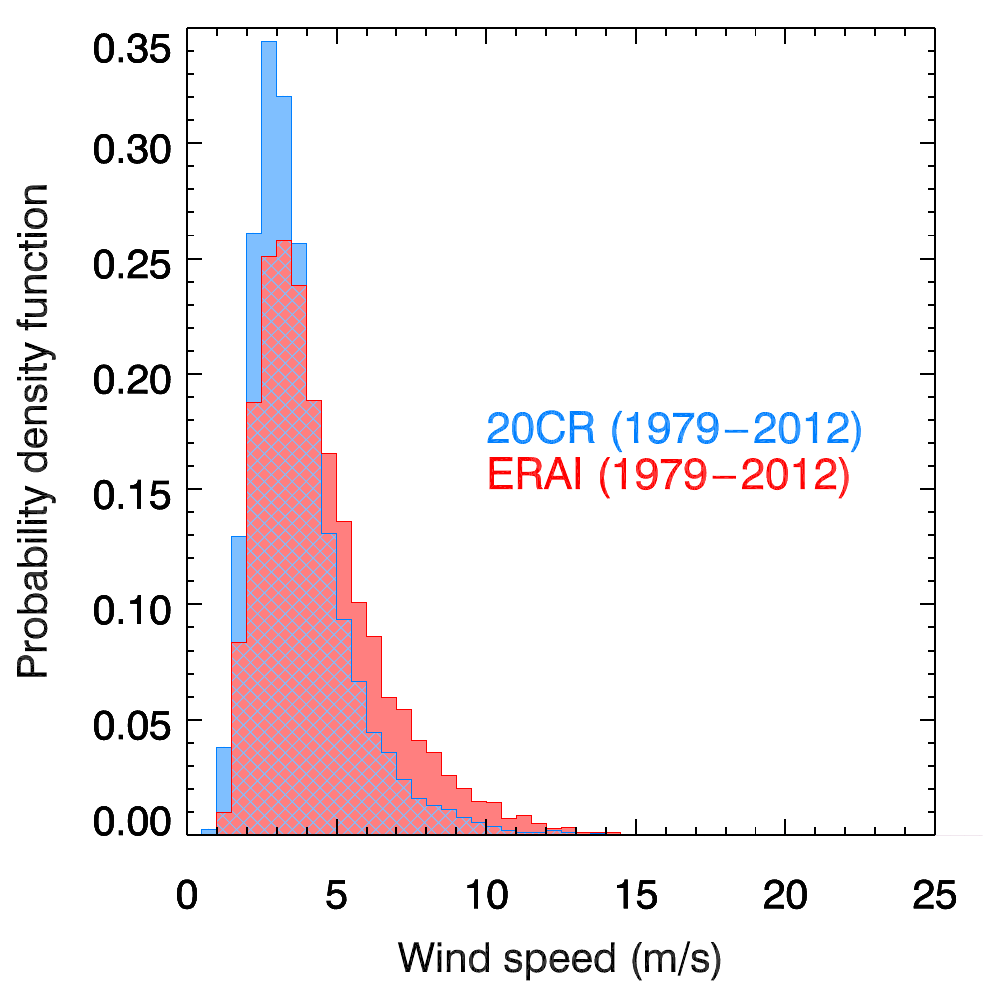} \includegraphics[width=\calibfourw]{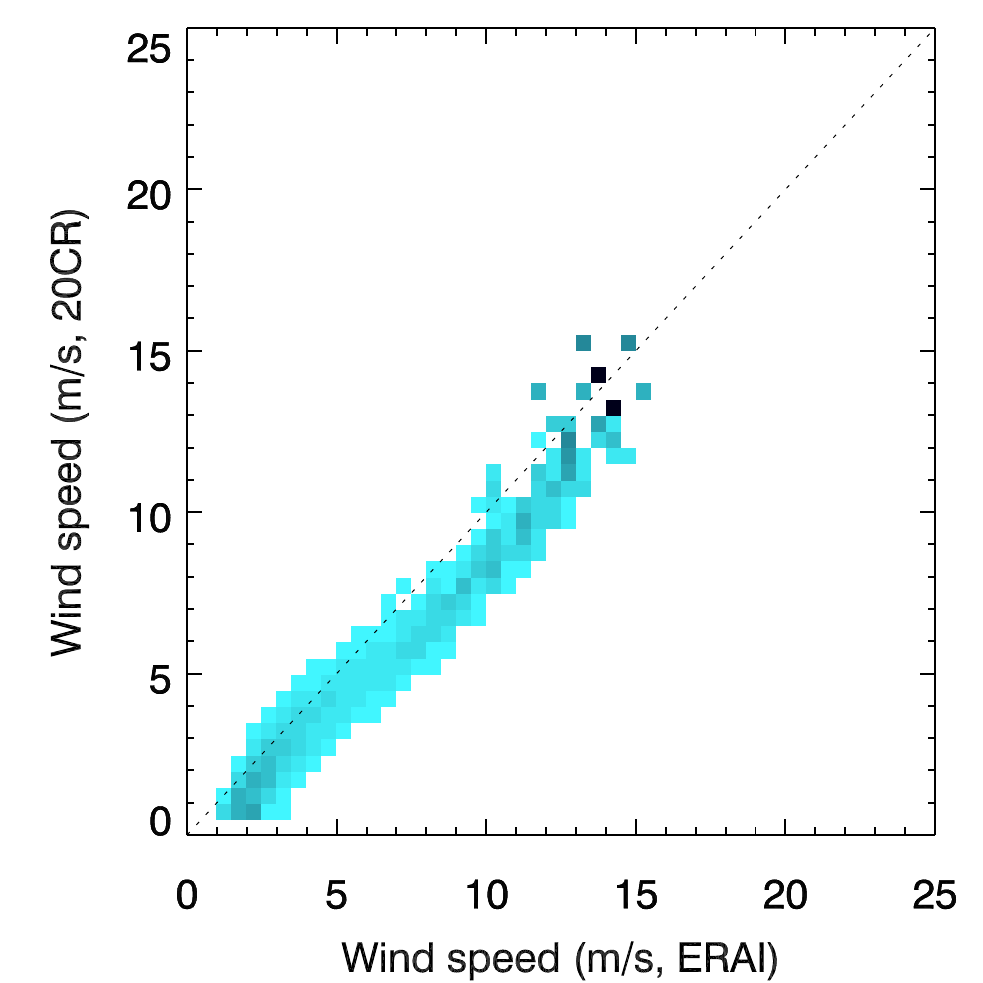}\\  \includegraphics[width=\calibfourw]{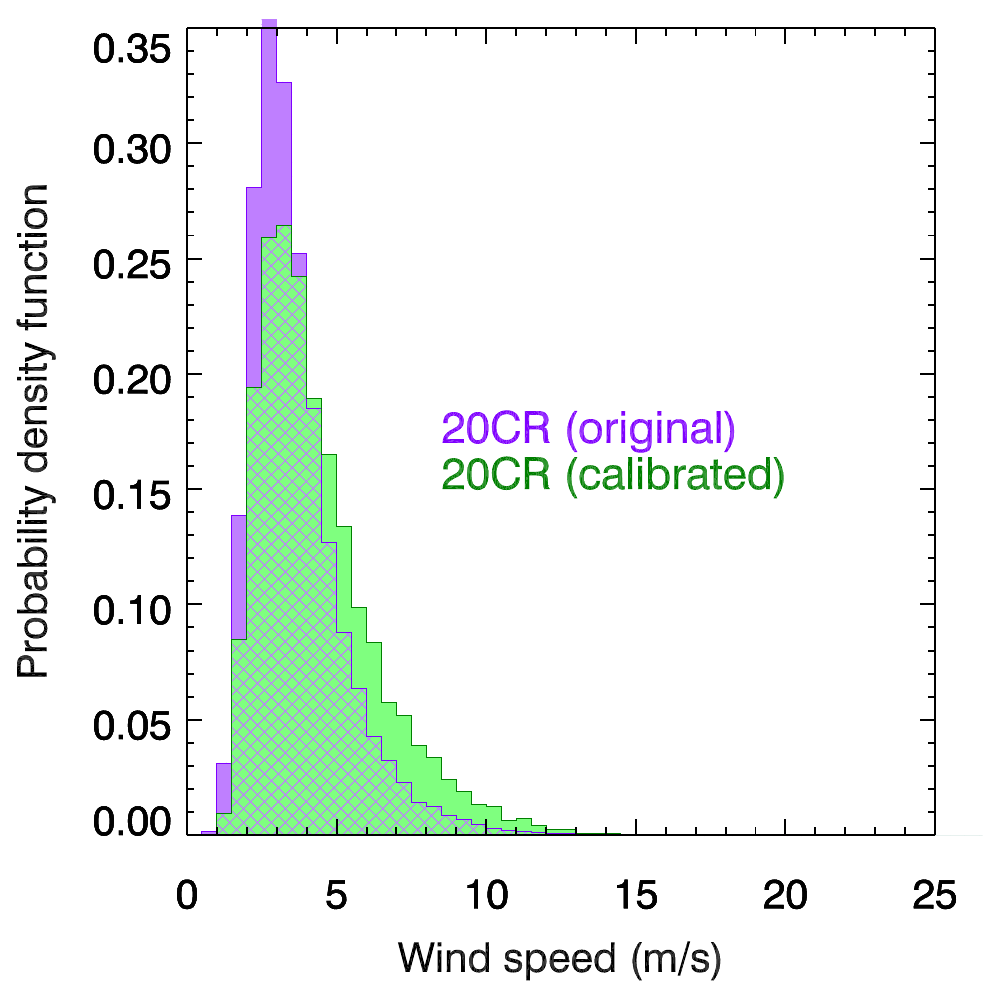} \includegraphics[width=\calibfourw]{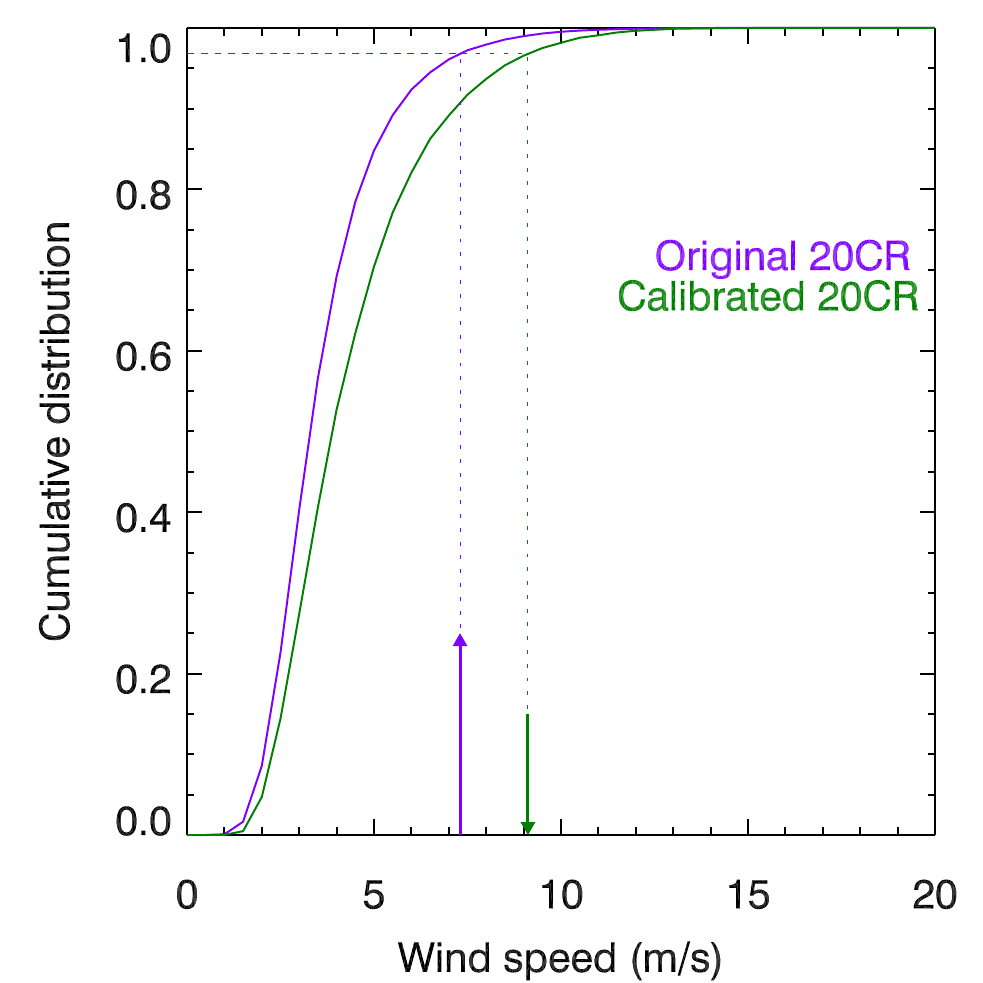}
\caption{Illustrating the calibration procedure, in terms of daily-mean wind speed probability distributions,  for the single grid cell covering north-western Germany (centre:  $8\degE$, $52\degN$); for the 20CR, the ensemble mean is used throughout for clarity.  Top-left: Daily-mean wind speed distributions for ERAI and 20CR over their intersecting time period.  Top-right: A visualisation of the conditional probability matrix $P(\Uerai_i | \Utwcr_j)$, such that each row $j$ is a PDF of ERAI wind speeds, given a particular 20CR wind speed $\Utwcr_j$ (i.e. the values along each row have the same sum).  Darker colours indicate higher frequencies in each ERAI PDF.  Bottom-left: Wind speed distributions over the full 142 years.   Bottom-right: The cumulative distribution function derived from the PDF histograms shown to the left.  The dotted lines and arrows illustrate the interpolation in the quantile-matching procedure used to convert wind speeds from the original 20CR data to their calibrated counterparts.  Here, a high wind speed for this cell from the original 20CR is transformed to its higher counterpart at the same level in the calibrated distribution.}
\label{f:calibmethod}
\end{figure}

It is important to note that the method we describe here is not unique.  Many other techniques for calibrating one data set with  another have been developed and used in climatological studies.  These are usually designed to compare reanalysis or model data with observations, or climate model data at different spatial scales, such as a global run with regional model output; see \cite{Teutschbein2012Bias}, \cite{Watanabe2012Intercomparison}, \cite{Lafon2013Bias} and references therein for recent reviews of methods.  
Compromises are reached between statistical complexity, data volumes,  direct numerical simulation, and time available.  In our case, we have chosen a relatively simple statistical procedure.

\subsection{Results of calibration procedure}
\begin{figure*}
  \centering
\includegraphics[width=\figbig]{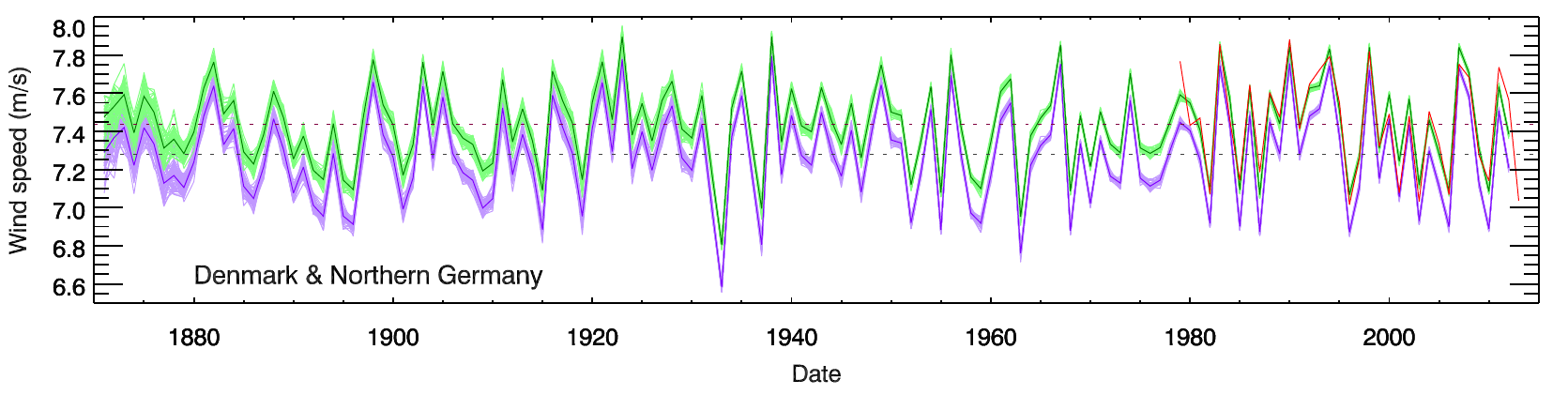} 
\caption{Time series of annual mean wind speeds for a region covering $9\degE$--$15\degE$ and $53\degN$--$57\degN$, showing the original 20CR data (purple), ERA-Interim (red) and calibrated 20CR (green).  For the 20CR data, the ensemble members are plotted in paler colours, with the ensemble means of the annual mean data plotted in darker colours.  Long-term averages are plotted as horizontal dashed lines.}
  \label{f:calibts}
\end{figure*}

Time series of annual mean wind speeds from both the original and calibrated   20CR data, and from ERA-Interim, are shown in Fig.~\ref{f:calibts} for a region covering Denmark and Northern Germany (using area-weighted averaging over the region).   The calibrated data retains the interannual variability of the original 20CR wind speeds, but with a climatology matching that of ERA-Interim over 1979--2012.

\begin{figure} \centering 
  \includegraphics[width=\figw]{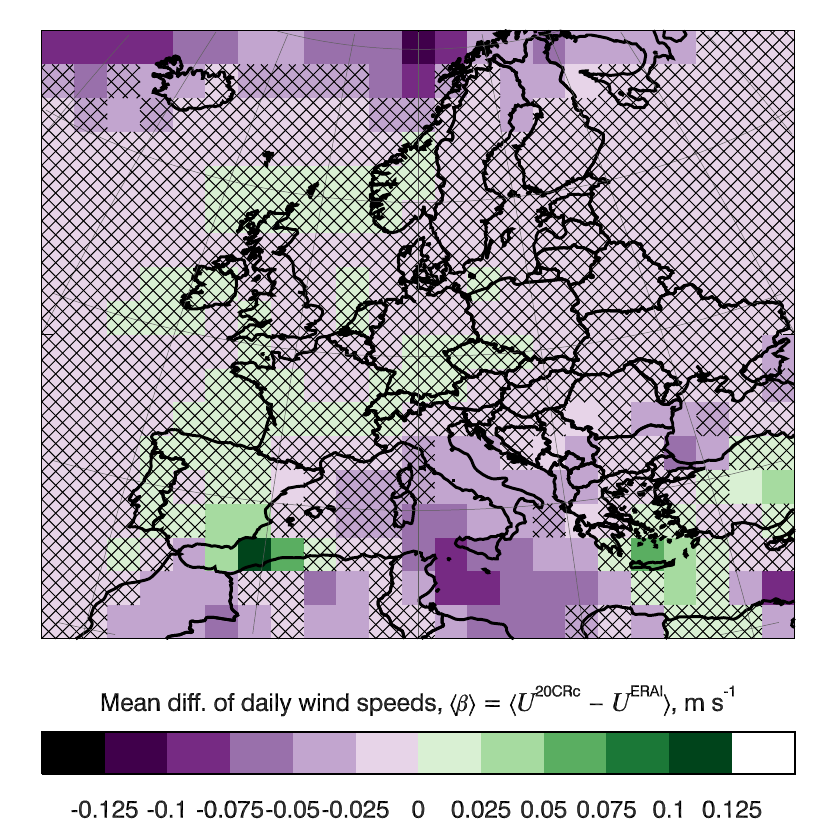} %{calib_meandiffmap_calib}
  \caption{Map of the remaining `bias' after calibration.  This can be compared to the map of the original bias in Fig.~\ref{f:meanbiasmaps}; note the colour scale covers much smaller values here.  Crosshatched areas are not significantly different from zero at the $1\%$ level, using a $t$-test.}
  \label{f:meanbiasmapsCalib}
\end{figure}

We map the bias remaining after our procedure in Fig.~\ref{f:meanbiasmapsCalib}.  This can be compared to the original bias maps in Fig.~\ref{f:meanbiasmaps} -- note that here the values are much smaller.    The mean bias $\langle \beta \rangle = \langle \Utwcrbc-\Uerai\rangle$ is consistent with zero almost everywhere (using a $t$-test at a $1\%$ significance level, as before).  An exception is a residual positive bias east of Gibraltar: we expect this area to be heavily affected by differences in how well the complex orography here is resolved between ERAI and 20CR. Two further exceptions occur in the central and eastern Mediterranean, which correspond to anomalies seen in other aspects of the 20CR data (see later sections), and which we discuss in more detail in Appendix~\ref{s:appx_ts}.

% They can be bigger as a column, but tidier as a row?
%  See pre-calib maps above.

The mean of the relative bias $\langle \beta_\mathrm{rel} \rangle$ (not shown) is $\leq 5\%$ almost everywhere.  Finally,  we note that the correlations between 20CR and ERAI after calibration (not shown) remain almost identical to those shown previously in Fig.~\ref{f:correlmaps}.

%===========================================================

%===========================================================

\section{Analysis and results}\label{s:results}
In this section we use the 20CRc data to analyse the distribution of wind speeds over Europe in various complementary ways.

%==============================================================================
\subsection{The European context: maps of the long-term average, variability and trends}\label{s:mapstrends}

\begin{figure}
\includegraphics[width=\figw]{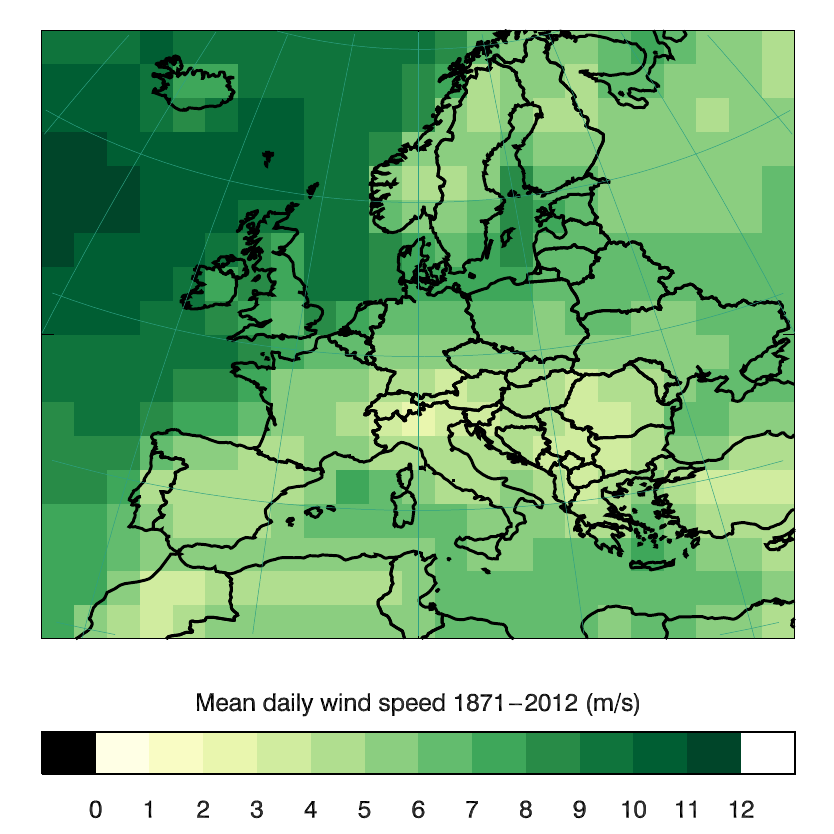} 
\caption{Long-term mean wind speed over Europe from the 20CRc data.}
\label{f:ltmapmean}
\end{figure}

The map of the 142-year mean wind speed in Fig.~\ref{f:ltmapmean} gives an overview of the  geographic distribution  of wind speeds over Europe. There is a noticeable land--sea contrast, although it is the mountainous regions that have the lowest mean wind speed, just as is seen in the uncorrected 20CR data \citep{BettThorntonClark2013}, and is inconsistent with observations.    This erroneous behaviour is a known consequence of the orographic drag schemes in atmospheric models \citep{Howard2007Correction}, and is particularly apparent when (as here) the orographic variability is on a much smaller horizontal scale than the model grid cells.  The spatial pattern in fact agrees very well with that derived by \cite{Kiss2008} from the $10\munit$ wind speeds covering 1958--2002 in the ERA-40 reanalysis \citep{Uppala2005}, although since they used winds at a lower level their mean values are correspondingly smaller.

It is important to note that the wind speeds shown here apply to the particular  spatial scale of this data set, which implies a certain amount of smoothing compared to values measured at a specific site.  For example, \cite{KirchnerBossi2013} use a complex statistical procedure to relate sea-level pressure from 20CR to wind speed observations at a range of meteorological stations in Spain.  Because they are statistically downscaling to this local scale, the mean wind speed they find is $2$--$3\windunit$ higher than we show in Fig.~\ref{f:ltmapmean}.

We map the wind variability in terms of its standard deviation.  The structure of the data set makes the calculation of the long-term standard deviation non-trivial: simply considering the ensemble-mean daily time series would result in a standard deviation that was negatively biased. Furthermore,  the ensemble members' time series are only continuous in 5-year chunks, and using them as if they were continuous throughout could potentially inflate the apparent variability at the discontinuities (although in practice the impact of this is likely to be very small).  To avoid such spurious signals and trends, we calculate the mean and standard deviation of daily wind speeds in each 5-year stream for each ensemble member, then take ensemble means for each period. We then combine these 5-yearly ensemble-mean standard deviations into single aggregate values for the full 142-year period, for each grid cell; see Appendix~\ref{s:varmethod} for details.

\begin{figure}
\centering
\includegraphics[width=\figw]{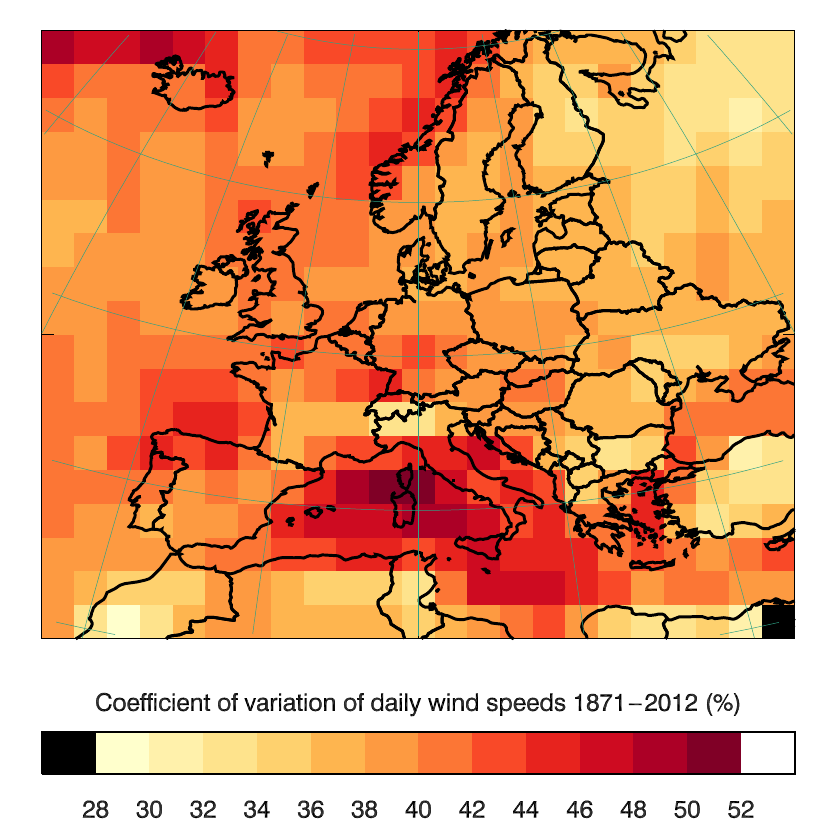} 
\caption{Map of the wind speed variability in terms of the coefficient of variation, i.e. the ratio of the standard deviation to the long-term mean.}
\label{f:varmap}
\end{figure}

Since the standard deviation of wind speeds tends to correlate with the mean, we show in Fig.~\ref{f:varmap} the wind variability in terms of the coefficient of variation,  the ratio of the standard deviation to the mean.   This shows that, in most areas, the wind speed standard deviation is $\sim 40\%$ of the mean.  The central Mediterranean has  proportionally higher variability, with Greece, Turkey and the Alps (whose orography will be extremely poorly represented) showing lower variability.

\begin{figure}
\includegraphics[width=\figw]{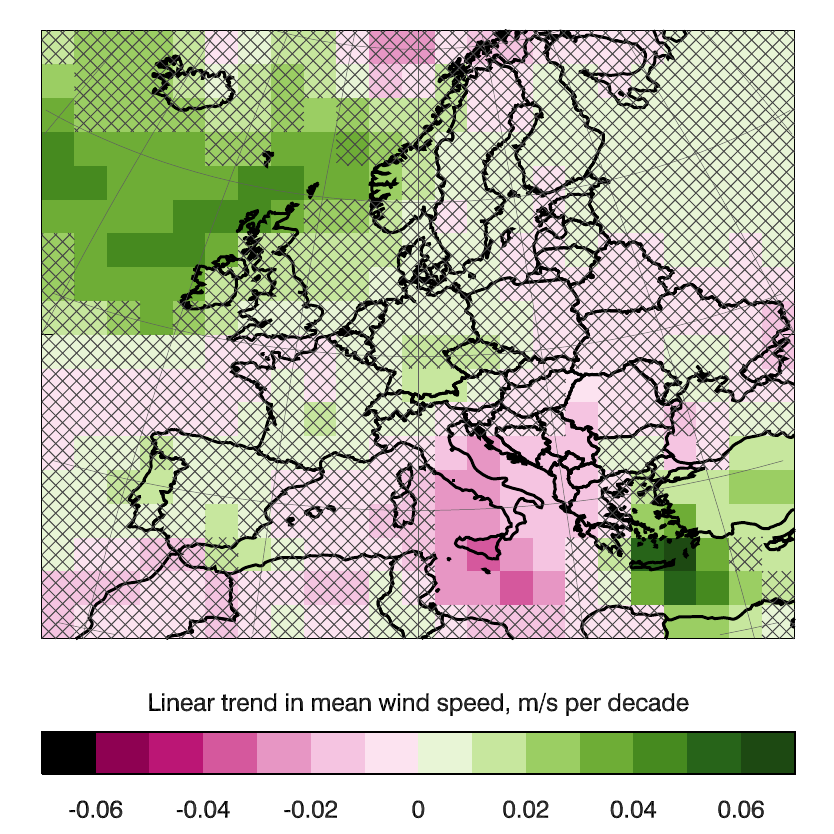} 
\includegraphics[width=\figw]{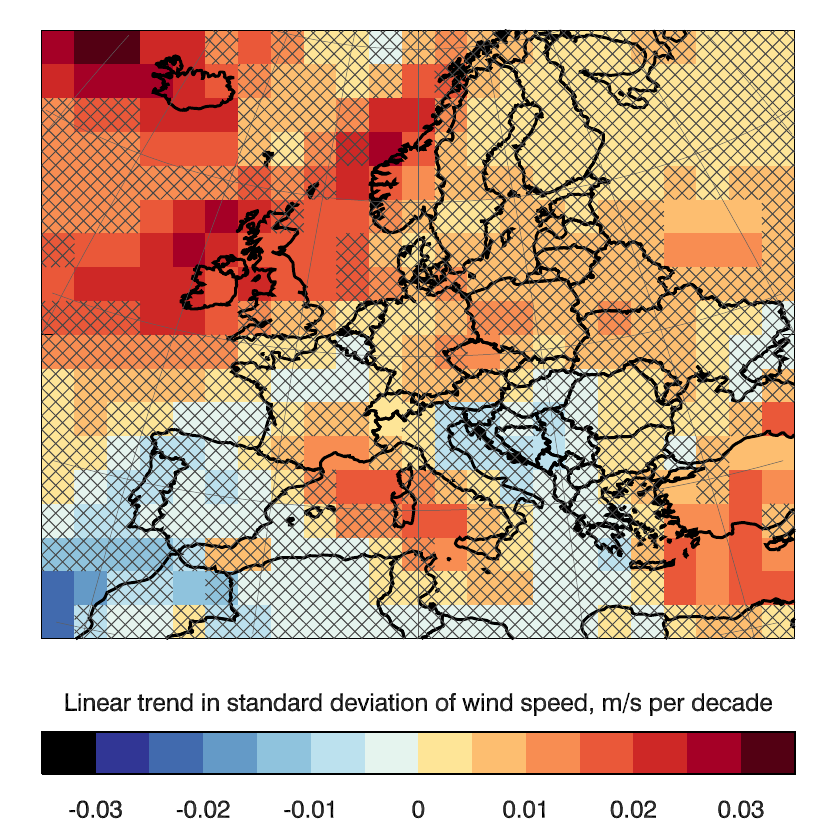} 
\caption{Map of the linear trend in the time series of ensemble-annual-mean wind speeds in each grid cell (top), and the ensemble-mean of the annual standard deviation of daily wind speeds (bottom), over 1871--2012. Crosshatched areas indicate where the trend is not significant at the $0.1\%$ level (see text for details).  } 
\label{f:trendmaps}
\end{figure}

The presence of any long-term trends in the mean or variability of wind speeds could have important consequences for wind farms, in terms of their future deployment, energy yield, and maintenance requirements.  Fig.~\ref{f:trendmaps} maps the trends in both the ensemble-mean annual mean wind speed and the ensemble-mean annual standard deviation of daily wind speeds.  The trends are found from the ensemble-mean annual time series using the Theil--Sen estimator \citep{theil1950rank,Sen1968Estimates}.  This is the median of the slopes between all pairs of points in the data set, and is more robust against outliers than simple linear regression, making it more suited to skew-distributed data such as wind speed. 

We test the significance of these trends at the $0.1\%$ level, using a Mann--Kendall test \citep{Mann1945,Kendall1975} modified using the method of \cite{Hamed1998Modified} to account for autocorrelation in the data \citep[following][]{Sousa2011Trends}; as is the case with much meteorological data, we expect adjacent timesteps to be correlated.  As with all significance tests, the result says whether the measured trend was unlikely, given the assumption of there being no true underlying physical trend.  If the probability of measuring the trend we did was below $0.1\%$, then we describe the trend as `significant', otherwise we regard it as consistent with zero.  We chose the particularly stringent threshold of $0.1\%$ to guard against detection of spurious trends; we only want to highlight trends we are \emph{very} sure are present in the data.

Some key points about long-term trends in European winds are immediately apparent from Fig.~\ref{f:trendmaps}.  Firstly, they are only on the order of a few centimetres per second per decade; and secondly, that in most areas of the continent, the trend is not significantly different from zero.  The trends in standard deviation show a similar spatial pattern, although at an even lower magnitude.

There are three areas of apparently  significant trend in annual wind speed that merit looking at in more detail: the areas of positive trend in the  Atlantic Ocean to the north and west of the British Isles, and the eastern Mediterranean around Crete; and the negative trend in an area of the central Mediterranean around the Italian peninsula and Sicily.   The Mediterranean regions were also anomalous in terms of their bias with respect to ERA-Interim (see previous section).  We look at the behaviour of wind speeds in these regions in more detail in Appendix~\ref{s:appx_ts}.

\cite{BettThorntonClark2013} used the same significance threshold for analysing trends in the uncorrected 20CR data, but measured trends using simple linear regression and $t$-tests to establish significance. While we consider the present technique to be more robust, the magnitude and spatial patterns of the trends are similar, and similar regions are highlighted as significant, pointing to genuine features in the underlying 20CR data.

As already discussed in the context of the mean wind speed, it is important to realise that these trends are those seen at the large scales of the 20CR data, and detailed physical or statistical modelling is required to downscale to a specific location.   Considering again the example of \cite{KirchnerBossi2013}, they find that the site in Spain they describe has a statistically significant negative trend in wind speed of around $-0.01\trendunit$.  In our results, the corresponding grid cell has a trend of around $+0.01\trendunit$, and is consistent with zero according to our test.

\subsection{Wind distribution time series}\label{s:windts}

\begin{figure}
\centering\includegraphics[width=\figw]{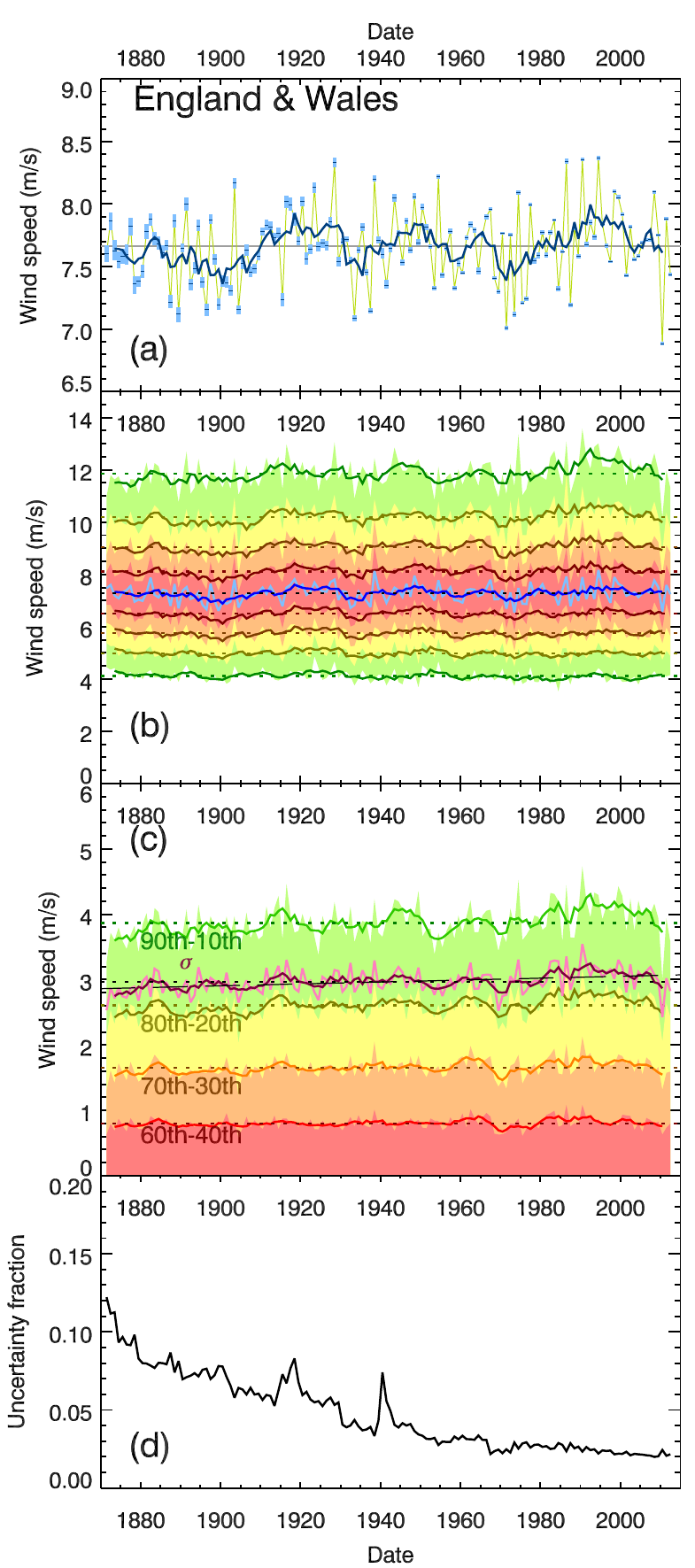} %{timeseries_altcolumnplot_englwal_dt1_calibtoERAI60m}
\caption{Time series of the wind speed distribution for a region covering $5\degW$--$1\degE$ and $51\degN$--$55\degN$.  In panels a--c, annual statistics are shown in light colours/shading, with darker lines showing the data smoothed with a 5-year boxcar window.  Panel a: Ensemble-mean annual mean wind speed. Individual years are shown with shading indicating the 10th/90th percentiles of the ensemble spread in the annual means. Panel b: Ensemble means of the deciles of the daily wind speed distribution each year (i.e. the 10th to 90th percentiles).  Panel c: Distribution half-widths, i.e. half the difference between symmetric decile pairs (as labelled); the standard deviation $\sigma$ is also plotted, with its trend shown as a thin black dashed line.  Panel d: The annual mean of the day-to-day standard deviation between ensemble members, as a fraction of the ensemble-mean annual mean wind speed. }
\label{f:regts_englwal}
\end{figure}

We use a region covering England \& Wales to to give an  example of how wind speed distributions can vary with time.  The time series of the area-averaged data from this region are shown in Fig.~\ref{f:regts_englwal}.  The annual mean wind speed (panel a) shows both large interannual variability, and (when smoothed with a 5-year boxcar window) strong decadal-scale variation.  For example, the smoothed series shows a clear increasing trend from around 1970 to a peak in the mid-1990s, followed by a return to near-average values after 2000.  When seen in the  142-year context however, these recent variations are not exceptional, and the year-to-year variability is always much greater.  Note that, for this region, the year 2010 is the extreme low-wind year.  This is linked to exceptionally cold months at the start and end of that calendar year, and a strongly negative NAO index in the 2009--2010 winter \citep{Cattiaux2010Winter, Osborn2011Winter, Brayshaw2012, Fereday2012Seasonal, Maidens2013Influence, Earl2013}.  The peak in wind speeds that occurs in the 1990s is another important feature in this region, and is also seen clearly in the observational record of wind speeds \citep{Earl2013}, in studies using geostrophic winds derived from  pressure observations \citep{Palutikof1992ukwind, Alexandersson2000Trends, Wang2009Trends}, and is consistent with the large positive NAO in these years \citep[e.g.][and references therein]{scaife2005}.  Indeed, much of the variability of wind speeds in this region is likely to be related to modes of climate variability such as the NAO and Atlantic Multidecadal Oscillation \citep[AMO, e.g.][]{Knight2006Climate}; further consideration of this requires careful seasonal breakdowns of both wind speed and these climate indices however, and is beyond the scope of this paper.

Our results bear a remarkable qualitative resemblance to those produced over 20 years ago by \cite{Palutikof1992ukwind} using geostrophic wind speeds (1881--1989) adjusted to match wind speed observations from a station in England over 1975--1984.   A key purpose of that study was to illustrate the long-term variability present in wind speeds, as it could have important implications for wind power production.  With the advent of larger datasets and greater computational capacity, we are able to re-emphasize their conclusions and consider the decadal-scale behaviour of the wind more robustly and in greater detail.

Considering the time series of the distribution as a whole (panel b), we can see that it follows the same decadal trends as the mean (panel a).  The distribution width (panel c) highlights that while the outer reaches of the distribution are subject to much variability, with the distribution width growing and shrinking over decades, the inner parts of the distribution are much more constant.  The standard deviation shown in that panel has a small but statistically significant positive trend, of $0.016\trendunit$.

Finally, the bottom panel shows the relative uncertainty in the data, in terms of the annual mean of the day-to-day ensemble spread.  As one looks further back, fewer observations are assimilated, and the ensemble members have more freedom to disagree with each other, resulting in increases in this measure of uncertainty.  Two peaks are present that are related to the reduction in data from Atlantic shipping during the World Wars; these spikes in uncertainty are ubiquitous for near-Atlantic regions.

In Appendix \ref{s:appx_ts}, we show similar plots for other regions that show particular features of interest, as already discussed.

Finally, we have given some consideration to the use of the \cite{weibull1951} distribution to concisely describe the wind speeds in our calibrated 20CR data.  However, as already mentioned, our use of daily average wind speeds means that Weibull distributions tend to provide a poor description of the data. Nevertheless, the Weibull scale parameter, which is proportional to the mean of the distribution, does tend to behave in the same way as the mean wind speeds in terms of variability and trends.  In particular, trends are of a similar magnitude and spatial pattern, and `anomalous' regions in the central and eastern Mediterranean, noted in previous sections, are also present.   Additional details and discussion are presented in the Supplementary Information.

%===================================================================

\section{Discussion and summary}\label{s:conclusions}
In this paper, we have demonstrated how century-scale reanalyses -- in particular, the  Twentieth Century Reanalysis, 20CR -- can be used for assessing the long-term trends and variability of near-surface wind speeds over Europe, through a calibration procedure to relate it to a higher-resolution satellite-era climatology (such as ERA-Interim), and subsequent careful analysis.

The long baseline of the 20CR means that it has great potential to inform wind speed assessments for the wind energy industry.  In general, reanalysis data is used in conjunction with dynamical and/or statistical downscaling techniques in order to reach the spatial scale of  wind farms, as part of the `model chain' in such assessments.  Often, it is the observation-rich and relatively high-resolution data sets of ERA-Interim and MERRA that provide that first reanalysis step. This limits any assessment of long-term variability, since they both only cover $\sim3$ decades.  By calibrating the 20CR data to match the climatology of ERA-Interim over their  period of overlap (1979--2012), this 142-year data set can be used in their place, providing a much more robust assessment of historic interannual and decadal variability in regions of Europe, and allowing the ``short-term'' trends of the past 10--30 years to be put into the longer-term context. 

\begin{figure}
\centering\includegraphics[width=\figw]{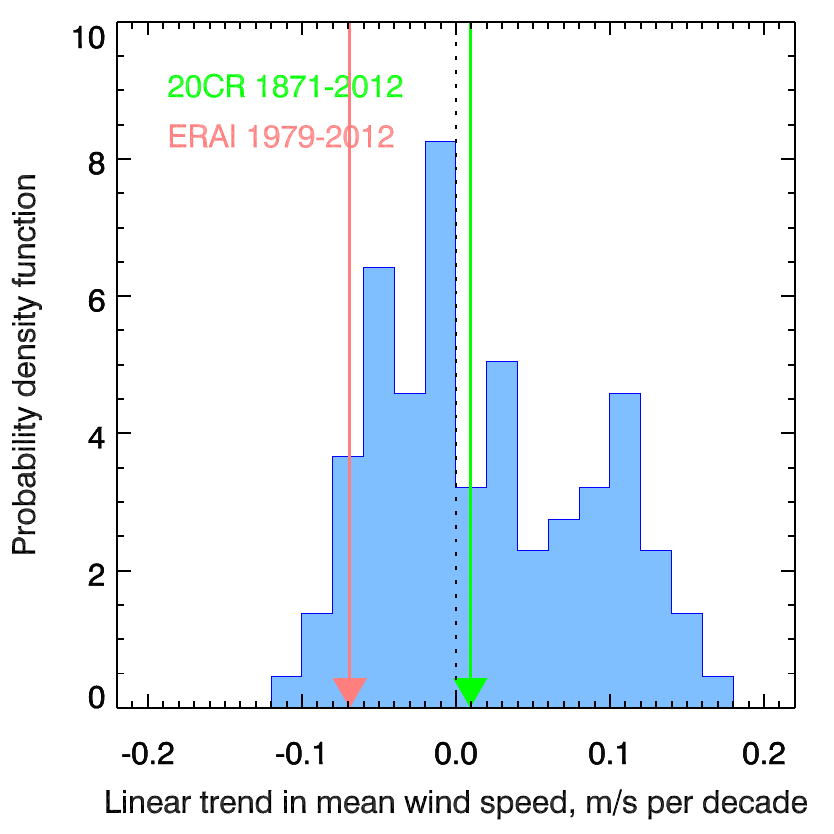} %{trenddistro_englwal_dt1_calibtoERAI60m}
\caption{Distribution of trends for the England \& Wales region.  The  34-year trends in annual-mean wind speeds from the calibrated 20CR data are shown as the blue histogram), overplotted with the full 142-year trend (green arrow).   The single 34-year trend from the overlapping period of ERA-Interim is shown as a red arrow.}
\label{f:trenddistro_englwal}
\end{figure}

To emphasise this point, we show in Fig.~\ref{f:trenddistro_englwal} the distribution of the 109 34-consecutive-year trends\footnote{i.e. the Thiel--Sen trend for 1871--1904 inclusive, and 1872--1905, 1873--1906, \ldots, and 1979--2012.} in annual mean wind speed for the England \& Wales region described in the previous section.  The full 1871--2012 trend is indicated and, as already shown, is near zero.  The trend from ERA-Interim for the 34 years of overlap is also marked, with a negative trend driven by the general reduction in wind speeds since the early 1990s.  It is clear that the strong multi-decadal variability in wind speeds means that attempting to estimate the long-term trend from a $\sim 30$-year sample can lead to misleading results.

The 20CR data is a rich source of information on the large decadal-scale variability of wind speeds.  However, it is not without limitations, and hence it does need to be analysed with care. For example, in areas of complex orography, near-surface wind speeds are strongly reduced at the spatial scale of the 20CR, making their variability more difficult to interpret.

As has been noted in other studies \citep{Compo2011, Bronnimann2012Extreme}, the ensemble nature of the 20CR needs to be taken into account when assessing long-term variability.   Disagreement between ensemble members can be large, especially in the early period of the data. This leads to the daily ensemble-mean time series having less variability than the individual members, and can cause apparent trends in variability over time. Therefore the daily ensemble-mean time series has little use in determining wind variability on long timescales, and the ensemble members should be used.

Assessment of trends over the full 142 years of the 20CR is complicated by the fact that the mid-point of the time series, and hence of a simple linear trend, is the 1940s.  The reduction in ocean-based measurements during both the First and Second World Wars causes spikes in uncertainty, and in some cases systematic spikes in the wind speeds themselves (see Appendix~\ref{s:appx_ts}).  Furthermore, the period after the  Second World War corresponds to a large increase in national and international programmes collecting greater amounts of weather data.  Taken together, the pre-1950s period is much more susceptible to greater  random and systematic uncertainties.  Measured trends in the 20CR data should therefore be treated with caution.  

We have shown in fact that all trends in 20CR surface wind speeds over Europe are either consistent with zero (in most locations), so small to be of little practical relevance (e.g. possibly in the North Atlantic), or due to systematic problems with the data (e.g. in the central and eastern Mediterranean and possibly the North Atlantic; see Appendix~\ref{s:appx_ts}).

It is clear that, for most wind energy applications, interannual variability and the large decadal-scale variability are more important than the very small long-term trends in historical  European wind speeds.  Using century-scale reanalyses such as the 20CR allows  wind resource assessment studies to incorporate more information on the historical decadal-scale variability at a site, which can reduce the uncertainties in the financial planning central to wind energy development.

%===========================================================================
\begin{acknowledgements}
PB would like to thank Adam Scaife, Chris Folland, Clive Wilson, Malcolm Lee, Jess Standen, Alasdair Skea and Doug McNeall, and the anonymous reviewer, for helpful comments and discussion.
Support for the Twentieth Century Reanalysis Project dataset is provided by the U.S. Department of Energy, Office of Science Innovative and Novel Computational Impact on Theory and Experiment (\href{http://www.er.doe.gov/ascr/incite/index.html}{DOE INCITE}) program, and Office of Biological and Environmental Research (\href{http://www.er.doe.gov/OBER/ober_top.html}{BER}), and by the National Oceanic and Atmospheric Administration \href{http://www.climate.noaa.gov/}{Climate Program Office}. 
ERA-Interim data was obtained from the ECMWF archive and are used under \href{http://data-portal.ecmwf.int/data/d/license/interim_full/}{license}.
\end{acknowledgements}

%===========================================================================

%===========================================================================
%=========================================================================

\appendix

\section{Choice of vertical level}\label{s:levels}
\begin{figure*}
  \centering
\includegraphics[width=\figbig]{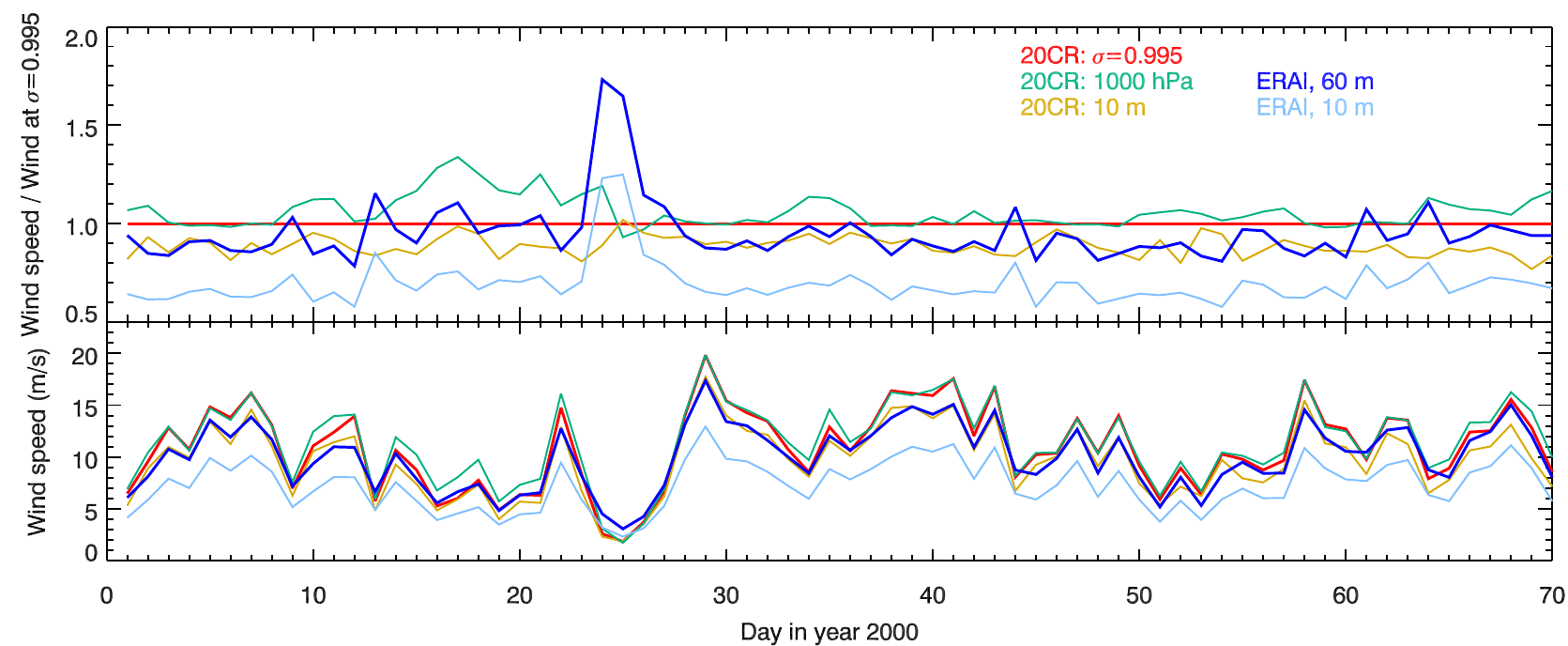} %{testlevels_ensmeanwindspeedts_both_englwal_2000}
  \caption{Demonstration of the daily mean wind speed at different near-surface levels, for the England \& Wales region.  Both panels show the results from the ensemble-mean 20CR winds at the $\sigma=0.995$ level, at $1000\hPaunit$, and at $10\munit$, as well as the $10\munit$ and $60\munit$ winds from ERA-Interim after regridding to match 20CR.  The top panel shows the daily-mean wind speeds as a ratio of the 20CR  $\sigma=0.995$ wind, and the bottom panel shows the actual wind speeds.}
  \label{f:leveldemo}
\end{figure*}

In Fig.~\ref{f:leveldemo} we compare the daily mean wind speeds from 20CR at the  $\sigma := P/P_\mathrm{surface} = 0.995$ level with those at the other available near-surface levels of $P=1000\hPaunit$ and $10\munit$, over an arbitrary period.  They have very similar variability behaviour, with the $1000\hPaunit$ winds tending to be slightly higher, and the $10\munit$ winds around $10\%$--$20\%$ lower.   

Fig.~\ref{f:leveldemo} also includes $10\munit$ and $60\munit$ winds from ERA-Interim.  The $60\munit$  vertical level was chosen as it is roughly similar to the height expected at $P=0.995P_\mathrm{surface}$.  A similar alternative would have been the $30\;\mathrm{m}$ model level, but we chose the higher level as it would be (marginally) less impacted by surface roughness;  $60\munit$ is also closer to wind turbine hub heights and thus more likely to be used for site-selection studies for the wind power industry.  Fig.~\ref{f:leveldemo} also suggests that the $60\munit$ winds provide a fairly good match to the 20CR $\sigma=0.995$ winds by eye.

%--------------------------------------------------------------------------

\section{Procedure for combining variances}\label{s:varmethod}
To avoid bias, we calculate variances of daily-mean wind speeds for each ensemble member separately, in consecutive $n$-year periods. In most cases, these periods are $n=5$ years, corresponding to the production streams of the 20CR (see Section~\ref{s:mapstrends}); the final period has $n=2$ years, covering 2011 and 2012.   These are combined into an aggregate population\footnote{We use population statistics here rather than sample statistics because we use data from \emph{every} day in each $n$-year period, rather than estimating the $n$-year standard deviation from a sample of days.} variance for the whole 142-year period over all ensemble members, using the following procedure.

If we consider a single time series of daily-mean wind speeds $U(t_j)$, at discrete timesteps labelled $j$, then we can divide it into a series of discrete $n$-year chunks labelled $i$, each containing $N_i$ days (leap years and the final 2-year period mean that not all $N_i$ are equal).

For each $n$-year period $i$, we can calculate the mean  $\overline{U}_i = N_i^{-1} \sum_j U(t_j)$, the mean of squares $\overline{U^2}_i = N_i^{-1} \sum_j U^2_j$,  and the variance $\sigma_i^2 = \overline{U^2}_i - \overline{U}_i^2$.  We store the mean and variance for each $n$-year period, for each ensemble member.

The aggregate means over all $n$-year periods (i.e. the 142-year means in our case) are simply
\begin{eqnarray}
  \label{e:aggmeans}
  \overline{U} &=& \frac{\sum_i N_i \overline{U}_i}{\sum_i N_i}, \\
  \overline{U^2} &=& \frac{\sum_i N_i \overline{U^2}_i}{\sum_i N_i}.
\end{eqnarray}
We can use these to write the aggregate population variance in terms of the mean and variance in each period:
\begin{eqnarray}
  \label{e:aggvarderivation}
  \sigma^2 &=& \overline{U^2} - \overline{U}^2\\
   &=&  \frac{\sum_i N_i \overline{U^2}_i}{\sum_i N_i} - \overline{U}^2\\
   &=&  \frac{\sum_i N_i \left( \sigma_i^2+\overline{U}_i^2 \right)}{\sum_i N_i} - \overline{U}^2. \label{e:aggvar}
 \end{eqnarray}

In practice, since we have stored the $n$-year means and variances for each ensemble member $m$, $\overline{U}_{i,m}$ and $\sigma^2_{i,m}$,   we take ensemble means to obtain $\overline{U}_i$ and $\sigma^2_i$ for each period.  These are then used to calculate $\overline{U}$ and $\sigma^2$ using equation~\ref{e:aggvar}.

%===========================================================================

\section{Additional regional time series}\label{s:appx_ts}
\begin{figure}\centering
\includegraphics[width=\figw]{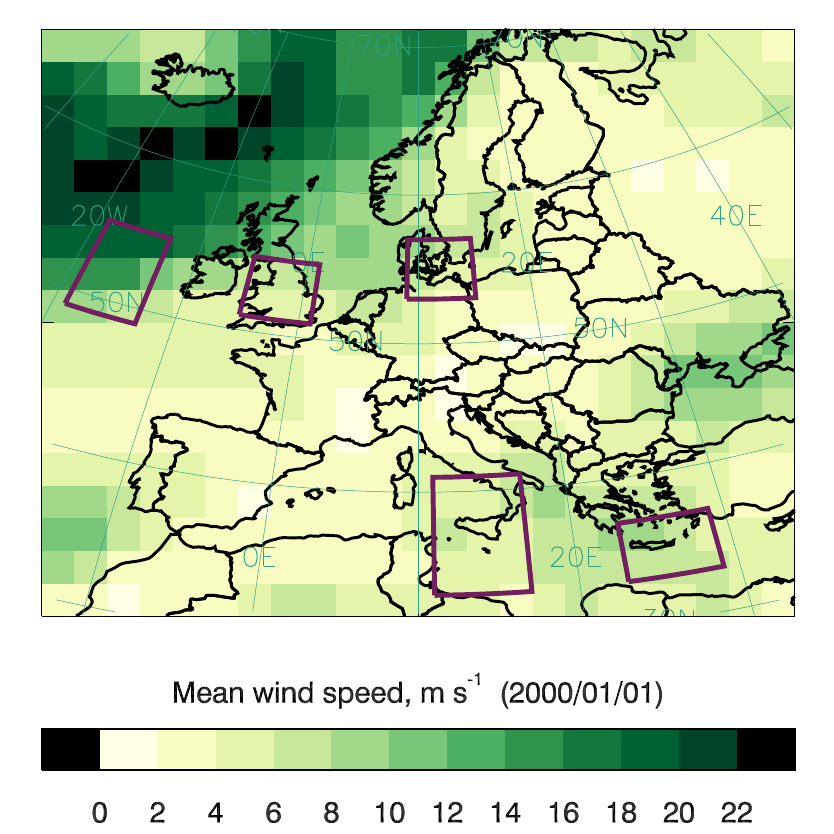} %{testplot_eurLAEA_regions_notitle}
\caption{Regions used in this study, overlaid on the mean wind speed from 20CR on an arbitrary day.  The regions defined in Table~\ref{t:regions} (on a regular lat--lon grid) are marked with boxes.  }
\label{f:regionsprojections}
\end{figure}

In this section we demonstrate the wind speed time series for some additional regions of interest, in the same manner as for the England \& Wales results discussed in Section~\ref{s:windts} (Fig.~\ref{f:regts_englwal}).  The regions are defined in Table~\ref{t:regions} and shown in Fig.~\ref{f:regionsprojections}, and were selected as areas of apparently `significant' trends in wind speed (see Fig.~\ref{f:trendmaps}).  As elsewhere in this paper, trends are calculated  using the Theil--Sen estimator, and their significance is tested using the modified Mann--Kendall test (see Section~\ref{s:mapstrends}).

\begin{table}
  \caption{Definitions of regions used in this study. Coordinates are given as ($\degr$ East, $\degr$ North).  Results for the first two regions are given in the main body of this paper, and this Appendix describes the bottom three regions.
  }
\label{t:regions}
\begin{tabular}{lr@{, }lr@{, }l}\hline
  Name     & \multicolumn{2}{l}{SW point}  & \multicolumn{2}{l}{NE point} \\\hline
  England \& Wales                & $ -5\degr$ & $51\degr$   &   $  1\degr$ & $55\degr$  \\
  Denmark \& Northern Germany     & $  9\degr$ & $53\degr$   &   $ 15\degr$ & $57\degr$  \\
  North Atlantic                  & $-19\degr$ & $49\degr$   &   $-13\degr$ & $55\degr$  \\
  Sicily \& Central Mediterranean & $ 11\degr$ & $33\degr$   &   $ 17\degr$ & $41\degr$  \\
  Crete  \& Eastern Mediterranean & $ 23\degr$ & $33\degr$   &   $ 29\degr$ & $37\degr$  \\\hline
\end{tabular}
\end{table}

% was originally 70mm each and figure*

\begin{figure} 
\centering\includegraphics[width=\figw]{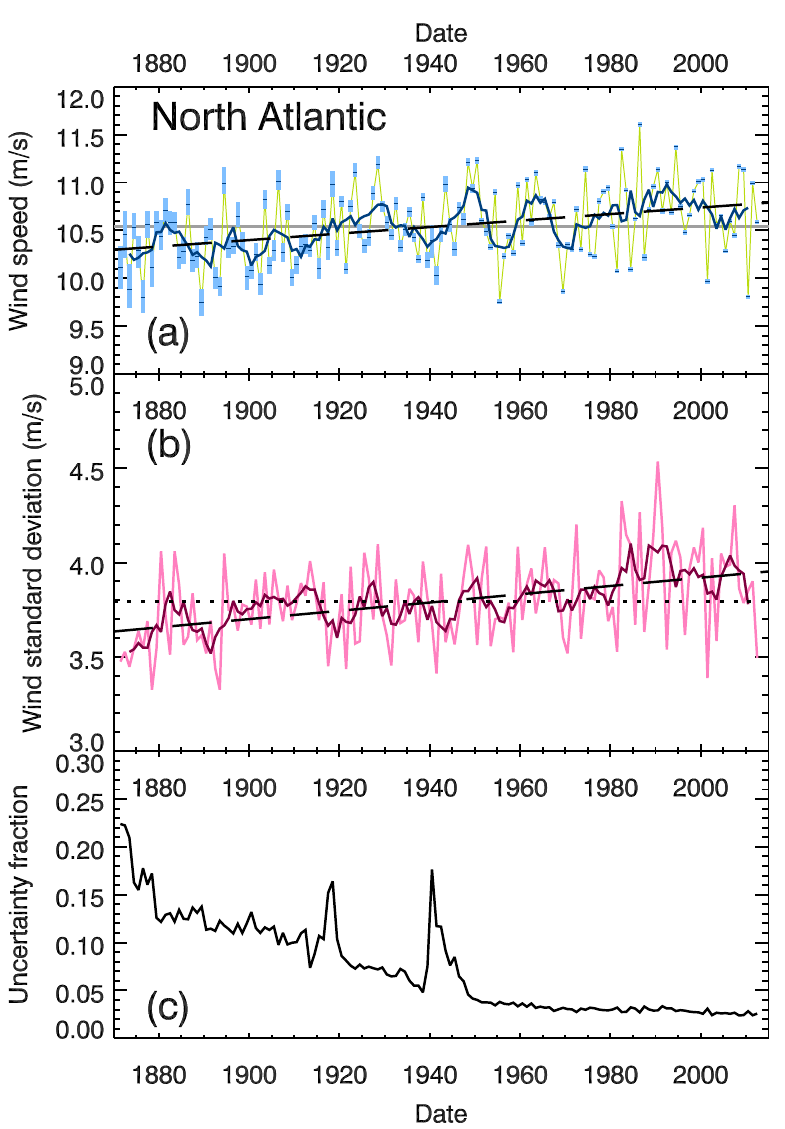} %{timeseries_columnplot_forappx_natlantic_dt1_calibtoERAI60m}
\caption{Time series of the wind speed distribution for the North Atlantic region, following Fig.~\ref{f:regts_englwal}.  The panels show annual values of mean wind speed (a), standard deviation of daily mean winds (b), and mean daily ensemble spread (c). Dark lines in (a) \& (b) give 5-year rolling averages, and trendlines are shown with black dashed lines; they are significant at the $0.1\%$ level (see text for details).  }
\label{f:regts_atl}
\end{figure}

Fig.~\ref{f:regts_atl} shows the results for the North Atlantic region.  As well as having much stronger and more variable wind speeds overall compared to England \& Wales, there are also significant positive trends in the annual mean wind speed and annual standard deviation of daily winds.  The increase in the uncertainty prior to the 1940s is much more striking than for the England \& Wales region, and casts a degree of suspicion on the trend in the annual mean wind speed.   It is plausible that  the apparent trend is simply due the winds prior to the 1940s in this location being systematically slightly lower than in the subsequent period, rather than being due to any true underlying physical mechanism.

A possible cause -- at least in part -- could be a difference between the variance in the observations ingested by the reanalysis, and the preferred variance of the underlying NWP model.  For example, if the observations are more variable than the model (e.g. if left running without assimilating data), then we might imagine that the 20CR data would have less  variance at early times when there are much fewer observations.  The skewed nature of wind speed distributions means that a trend in variance could lead to a trend in mean wind speeds too.  However, the 20CR employs a covariance inflation process  (see \citealt{Compo2011} and references therein for details), which will act in the opposite direction.  Without further detailed study of the model behaviour, these ideas remain at the level of speculation.

The WASWind data set produced by \cite{Tokinaga2011Wave}, based on ship-based measurements of wind and wave heights, has a negative trend in winds for the North Atlantic over 1950--2008.   In our data, the trend over the 1950--2010 period is positive, but not significantly different from zero.   The weakness of both trends, and difficulties with the observations in both cases, means that it is hard to be conclusive about the `true' situation.

However, the negative trend we see between around 1990 to around 2005 \emph{is} seen in the WASwind data, and \cite{Vautard2010} have shown that it is also present in the ERA-Interim data.  Finally, \cite{Vautard2010} found a negligible trend in the North Atlantic in the NCEP/NCAR Reanalysis \citep{Kalnay1996NCEPNCAR} over 1979--2008, which is also be consistent with our results.

Long-term trends in extreme wind speeds and storminess in the North Atlantic have been discussed in \cite{Wang2009Trends, Wang2011Trends, Wang2013Trends}, \cite{Krueger2013Inconsistencies, Krueger2014Comment}, and \cite{Wang2014Storminess}.  These studies relate extreme winds derived from long-term pressure records with those derived from the 20CR data set, and demonstrate both the decadal-scale variability that we see here, and the difficulty of drawing definitive conclusions from trend analysis with this data: different analysis methods can produce very different results, and the 20CR data prior to the 1950s should be treated both carefully and sceptically.

\begin{figure} 
\centering\includegraphics[width=\figw]{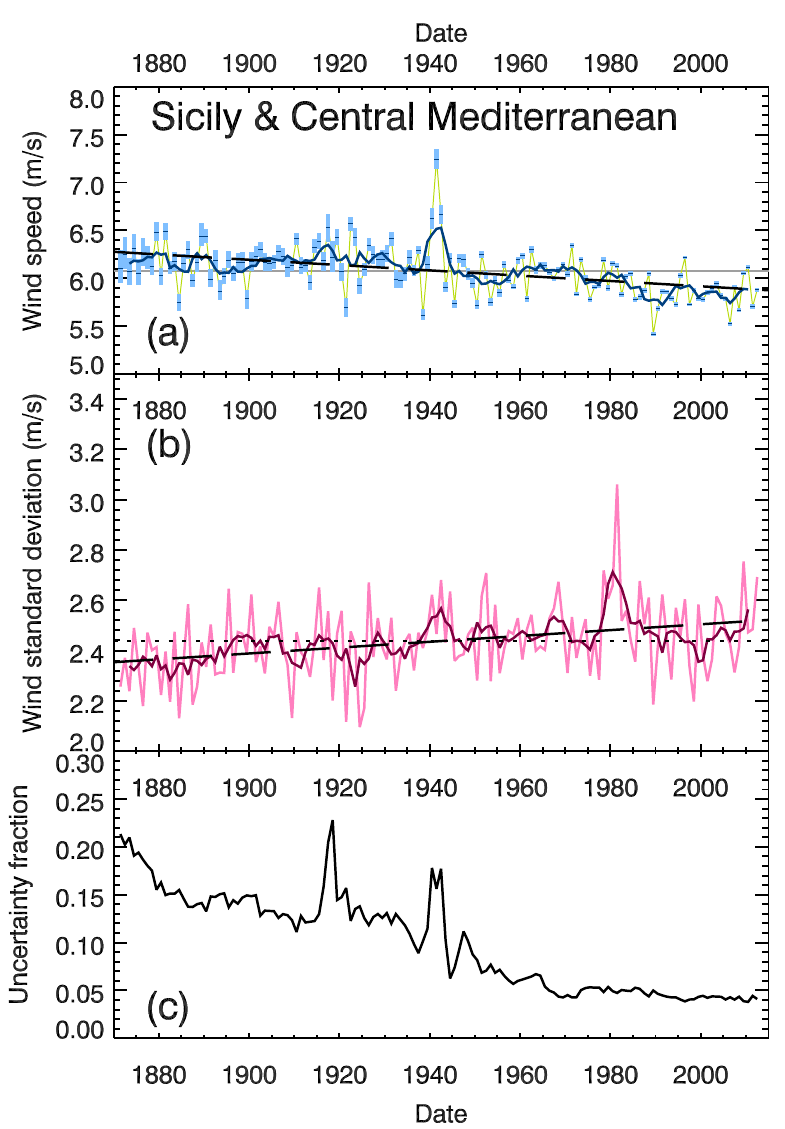} %{timeseries_columnplot_forappx_medcsicily_dt1_calibtoERAI60m}
\caption{As Fig.~\ref{f:regts_atl} but for  the wind speed distribution in the Sicily \& Central Mediterranean region. While the annual mean wind speeds have a significant negative trend (black dashed line in panel a, see text for details), there is no significant trend in the standard deviation (panel b). }
\label{f:regts_medcent}
\end{figure}

The Sicily \& Central Mediterranean region appeared to have a significant negative trend in wind speeds in Fig.~\ref{f:trendmaps}; the time series for the annual mean wind speeds in that region is shown in Fig.~\ref{f:regts_medcent}.   We can see again the high levels of uncertainty prior to the 1950s, and a particularly anomalous spike in wind speeds around 1940--1942.  If we take that spike to be indicative of the kind of systematic errors that might be present in the early half of the data, but not captured by the ensemble spread, then it is not unreasonable to suppose that the entire period prior to the 1940s could be showing higher wind speeds than it should, and thus accentuating a negative trend.

However, there does appear to be a more genuine negative trend in the data from the 1950s onwards, where the uncertainties are much more reasonable.  We find that the Theil--Sen slope for the 1950--2012 period is very similar to that of the full 142 years, although in this case it is \emph{not} significantly different from zero at the $0.1\%$ level.  However, as there are so few decades available from the 1950s, it is difficult to know how such an apparent trend relates to the decadal-scale oscillations that we see here, and in other regions.

Overall, the uncertainties in the data make it extremely difficult to separate decadal climate variability, systematic errors, and genuine long-term trend.  

\cite{Pirazzoli2003Recent} looked at trends in the observed wind speeds over a similar region using station data mostly covering 1951--2000.  They found a mixture of trend behaviours: most stations showed a negative trend prior to the 1970s that then became positive; some stations showed no trend, or trends which became negative from the 1970s onwards.  In our data, which will not be able to resolve the complex coastal and orographic features of the region, we can see that the 5-year running mean appears to be increasing from the 1950s, changing to a negative trend after the 1970s.  While this clearly disagrees with the \cite{Pirazzoli2003Recent} results from some stations, it is unclear how the variety of different observed behaviours in this complex terrain should combine to produce an aggregate  trend on the large scales of the 20CR.  In any case, the trends in the 20CR data are extremely slight; the main conclusion from our data should be that interannual variability is vastly more important than any trend for this region over a period as short as 50 years.

\begin{figure} 
\centering\includegraphics[width=\figw]{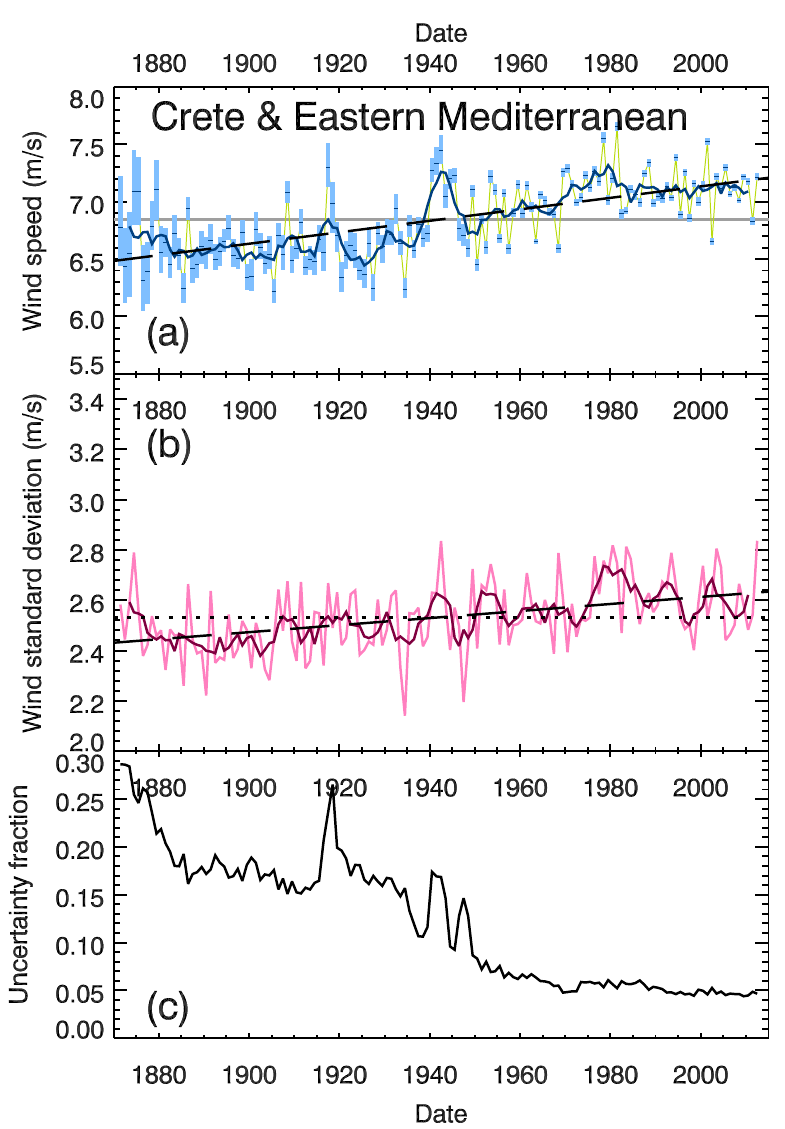} %{timeseries_columnplot_forappx_medecrete_dt1_calibtoERAI60m}
\caption{As Fig.~\ref{f:regts_atl} but for the wind speed distribution for the Crete \& Eastern Mediterranean region.  Both the annual mean wind speeds (panel a) and the standard deviations (panel b) have statistically significant trends, marked as black dashed lines; see text for details.  }
\label{f:regts_medeast}
\end{figure}

Finally, we show the time series for the  Crete \& Eastern Mediterranean region in Fig.~\ref{f:regts_medeast}.  In this case, the apparent overall trend is positive.   There is again a spike in wind speeds in the early 1940s, and a suggestion that the data prior to the 1950s could be systematically shifted relative to the latter period.  Another interesting feature is that the early period until around the 1920s shows a slight decrease over time; if we exclude the 1940s spike, this then appears to be followed by a long generally-increasing period until the 1980s, after which the wind speeds have been relatively constant.  

As before, the uncertainties in the data, both systematic and as seen in the ensemble spread, coupled with the expectation of decadal-scale variability and a time series that is ``only'' 14 decades long, mean that it is impossible to know from this data alone how ``real'' such a very long oscillation might be.  If we allow for systematic shifts in the 1940s and before, the data is consistent with there being no long-term trend, but with decadal-scale variations underlying large interannual variability, as in other regions.  What we can say with some certainty however is that the wind speeds in this region have been higher since the 1970s than they were in the 1950s and 1960s -- with the caveat of there being strong interannual variability.

%===================================================================
%===================================================================

%\bibliographystyle{spbasic}    % basic style, author-year citations
%\bibliography{ltwind_distro}   % name your BibTeX data base
%===================================================================
%===================================================================

%=========================================================================
%=========================================================================
\clearpage
\section{Supplementary Information: Weibull distribution fits}\label{si:weibull}
% To be inserted into the paper as a supplementary appendix (e.g. for arXiv)
% or to be produced as a separate supplementary document (for the journal)

%=============================================================================

Concise measures of the wind speed distribution are convenient when comparing different locations and periods.   Wind speed distributions are frequently analysed in terms of the \cite{weibull1951} distribution.  The  probability density function (PDF) for a Weibull-distributed random variable $U \sim \mathcal{W}(\lambda,k)$ is given by
\begin{equation}\label{e:weibullpdf}
 P(U|\lambda,k) = \frac{k}{\lambda} \left(\frac{U}{\lambda}\right)^{k-1} 
         \exp{ \left[ -\left(\frac{U}{\lambda}\right)^k \right] },
\end{equation}
which has two free parameters: the scale $\lambda$, proportional to the mean of the distribution, and the dimensionless shape ($k$), which determines the skewness.  The distribution width and peak location are related to both parameters. 

The choice of the Weibull distribution for wind speed analysis has a partly theoretical and partly pragmatic basis \citep{Hennessey1977Some}.  From a theoretical standpoint, the simplest case is the (one-parameter) Rayleigh distribution, which describes the magnitudes of two-component vectors when the components are uncorrelated and Gaussian-distributed, with equal variance and zero mean.  A Rayleigh-distributed random variable $U \sim \mathcal{R}(\varsigma)$ has a PDF given by
\begin{equation}\label{e:rayleighpdf}
  P(U|\varsigma) = \frac{U}{\varsigma^2} \exp{\left[-\frac{U^2}{2\varsigma^2}\right]},
\end{equation}
in terms of its scale parameter $\varsigma$. 

Describing real wind speeds requires the more general case of correlated components with unequal variances.  While this does not itself have a closed-form mathematical expression, the Weibull distribution in fact provides a very good approximation to this more general distribution \citep{Harris2014Parent}. From the definitions above, it can be seen that the  Rayleigh distribution can be viewed as a special case of a Weibull distribution with $k=2$ and $\lambda=\varsigma\sqrt{2}$.  Empirically, the Weibull  distribution has been found to fit reasonably well with observed distributions of wind speeds for many decades,  and it aids comparison with previous work to continue using it (although not uncritically).\footnote{Over the past four decades, many studies have been (and continue to be, e.g. \citealt{Baile2011MRice, Qin2012Two, Morrissey2012Tractable,  2012arXiv1211.3853D, Monahan2012Can}) published on using other functions to better fit the distribution of wind speeds, balancing simplicity against adding more free parameters (for example generalising the Weibull distribution further into the generalised gamma distribution); see the review of \cite{Carta2009Review}, and references therein.}

We are using daily-mean wind speeds in this study, and this averaging has an impact on the wind speed distribution.  Taking a theoretical example, consider a series of wind vectors whose magnitudes follow a Rayleigh distribution, i.e. which have  Gaussian-distributed ($\mathcal{N}(\mu,\sigma)$) components with variance $\sigma_u^2$:
\begin{eqnarray}
  \vec{U} &=& \left(\begin{array}{c}
      u \sim \mathcal{N}(0,\sigma_u)\\
      v \sim \mathcal{N}(0,\sigma_u)\\
    \end{array}\right) \label{e:rayleighvecsample}\\
  U \equiv |\vec{U}| &\sim& \mathcal{R}(\sigma_u) \sim \mathcal{W}(\sigma_u\sqrt{2},2) 
\end{eqnarray}
We can take averages over every $n$ points in this series: for example, if we imagine the data represent ``6-hourly'' wind speeds, then taking averages of every $n=4$ points would give us ``daily'' mean wind speeds.  If we consider the resulting mean wind vectors $\langle\vec{U}\rangle$, they  remain Rayleigh distributed:
\begin{eqnarray}
  \langle v\rangle \sim \langle u \rangle &=& \frac{1}{n}\sum u
     \sim \mathcal{N}\left(0,\sigma_u\frac{\sqrt{n}}{n}\right)\\
  \langle\vec{U}\rangle &=& \left(\begin{array}{c}
      \langle u\rangle\\
      \langle v\rangle\\
    \end{array}\right)
  \sim \left(\begin{array}{c}
      \mathcal{N}(0,\sigma_u/\sqrt{n})\\
      \mathcal{N}(0,\sigma_u/\sqrt{n})\\
    \end{array}\right) \\
  |\langle\vec{U}\rangle| &\sim& \mathcal{R}(\sigma_u/\sqrt{n}) \sim \mathcal{W}(\sigma_u\sqrt{2/n},2) 
\end{eqnarray}
The mean and standard deviation of the resulting distribution are both reduced by $1/\sqrt{n}$ compared to those of the underlying wind speeds.

If however we consider the ``daily'' averages of the wind speed magnitudes, $\langle|\vec{U}|\rangle$, then the result is no longer Rayleigh-distributed. It retains the same mean value of the underlying distribution, although the standard deviation is again reduced by $1/\sqrt{n}$. 

\begin{figure}[t]
\includegraphics[width=\figw]{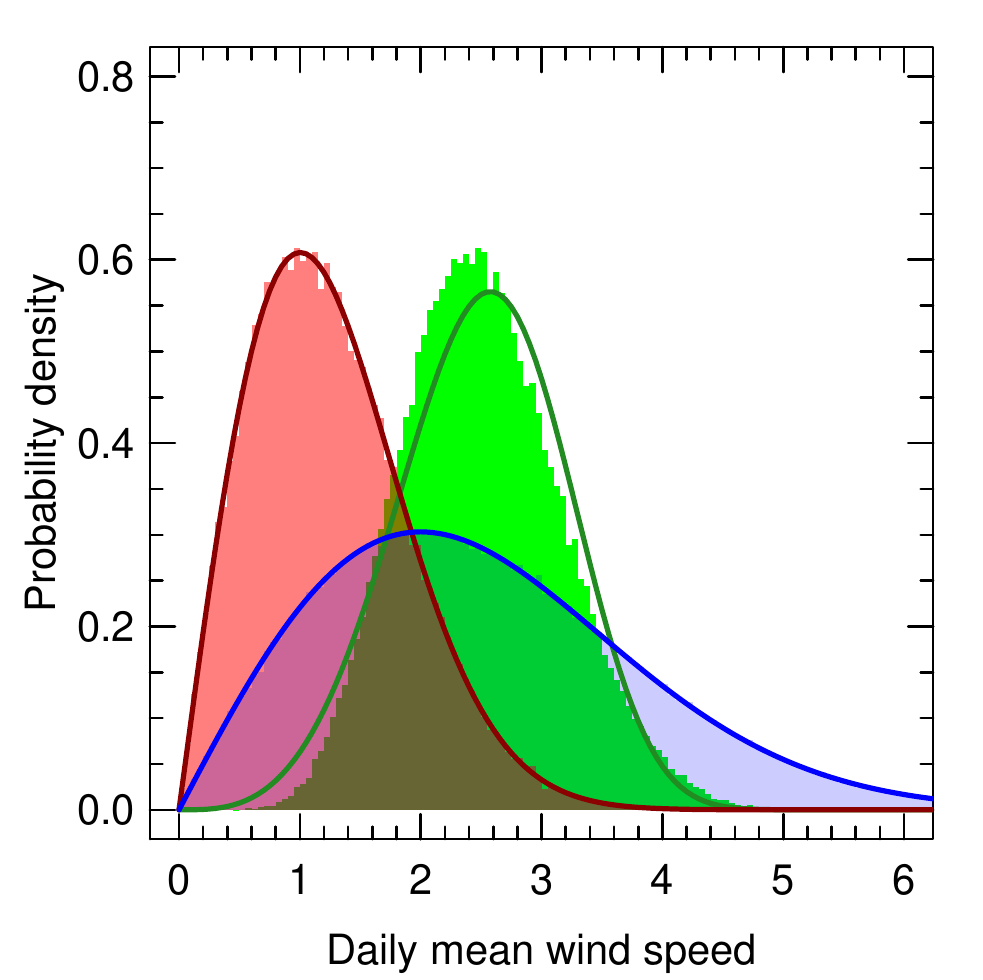}
\caption{Illustration of the impact of time-averaging on Rayleigh-distributed wind speeds. PDFs are shown as shaded histograms, with best-fitting Weibull distributions shown as solid lines.  We use 400\,000 random samples of $\vec{U}=(u,v)$ following equation~\ref{e:rayleighvecsample} with $\sigma_u=2$; the distribution of $|\vec{U}|$ is shown as the blue PDF.  The other histograms show the result of  averaging every $n=4$ samples (as if we are taking daily averages of 6-hourly wind speeds), resulting in 100\,000 data points.  The red histogram shows the distribution of magnitudes of mean wind vectors, $|\langle\vec{U}\rangle|$; this remains Rayleigh-distributed, with a reduced overall average value. The green histogram to the right shows the distribution of the means of wind speed magnitudes, $\langle|\vec{U}|\rangle$; this is no longer Rayleigh- (or Weibull)-distributed, but retains the overall mean value of the original distribution. }
\label{f:rayleighdemo}
\end{figure}

These results are summarised graphically in Figure~\ref{f:rayleighdemo}.  The key point is that, while we usually want the daily averages of wind speed magnitudes, this will necessarily pull the tails of the distribution in towards the mean value, making it more Gaussian and less Weibull-like\footnote{A similar effect will be present when area-averaging data between different spatial grids.}.

The impact of the distribution taking on this shape is that, while we expect the best-fitting  Weibull parameters to still respond systematically to variations in the wind speed distribution, the distributions that one would derive from those Weibull parameters are not accurate representations of the wind speeds.  For example, we expect that quantiles of the wind speed distribution estimated using the Weibull fits would be highly inaccurate; however, the relative movements over time of the scale parameter at least will broadly describe shifts in the peak location of the wind speed distribution.

Using the calibrated 20CR data, we calculate maximum-likelihood estimates of the Weibull parameters \citep{Carta2009Review}\footnote{Weibull  estimation is performed in practice using the \texttt{fitdistr} function in the MASS package for R \citep{MASSbook}.}  for each ensemble member in each of the 28 consecutive 5-year periods in 1871--2010 in each grid cell, using the distribution of daily-mean wind speeds (we omit 2011 and 2012 so that each time period uses the same number of years).  We do the same for the area-averaged wind speed ensemble member time series in  the standard analyses regions used in the main paper.

The goodness of fit of each Weibull distribution is assessed using a chi-square test: in each case we compare the histogram of the daily-mean wind speed data with that derived from the best-fitting Weibull PDF, using wind speed bins\footnote{Following common practice \citep[after][]{Cochran1954Some}, bins where the Weibull frequencies are $<5$ are merged with adjacent categories.} of $1\windunit$ and a significance level of $1\%$.  Passing the goodness-of-fit test implies that the data are consistent with  being drawn from the given Weibull distribution, at the $1\%$ significance level; it is a viable Weibull.

\begin{figure}
\includegraphics[width=\figw]{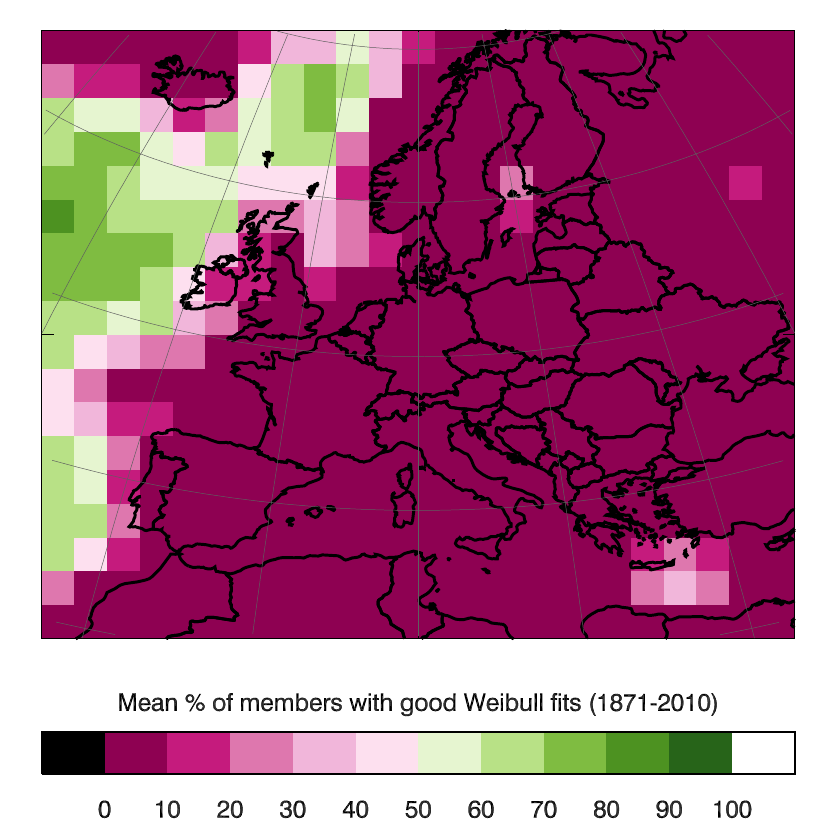}
\caption{Goodness of fit map. Each point shows the percentage of ensemble members whose Weibull fits pass the $\chi^2$ test (see text), averaged over the 28 5-year time periods. }

\label{f:weibgofmap}
\end{figure}

Figure~\ref{f:weibgofmap} shows a map of the mean, over the 28 5-year periods, of the percentage of ensemble members whose Weibull fits pass the $\chi^2$ goodness-of-fit test at the $1\%$ significance level.  While there appears to be some distinction between the better fits in the ocean and poorer fits in  continental regions, this is not as simple as a land--sea contrast.  For most locations in our domain, very few of the fits are `good'.

To demonstrate that these poor fits are not a product of the re-gridding or  calibration procedures we have applied to the data, we show  in Figure~\ref{f:weibgofpdfs} the PDFs of daily wind speeds at a particular grid cell location (western England) for a single 5-year period, comparing the different data sets we use.  While area-averaging the ERA-Interim data certainly exacerbates the problem, moving data in the tails towards the mean, neither the original ERAI or 20CR data have very good Weibull fits.

\newcommand{\figwhere}{0.24\textwidth} 
\begin{figure*}\centering
\includegraphics[width=\figwhere]{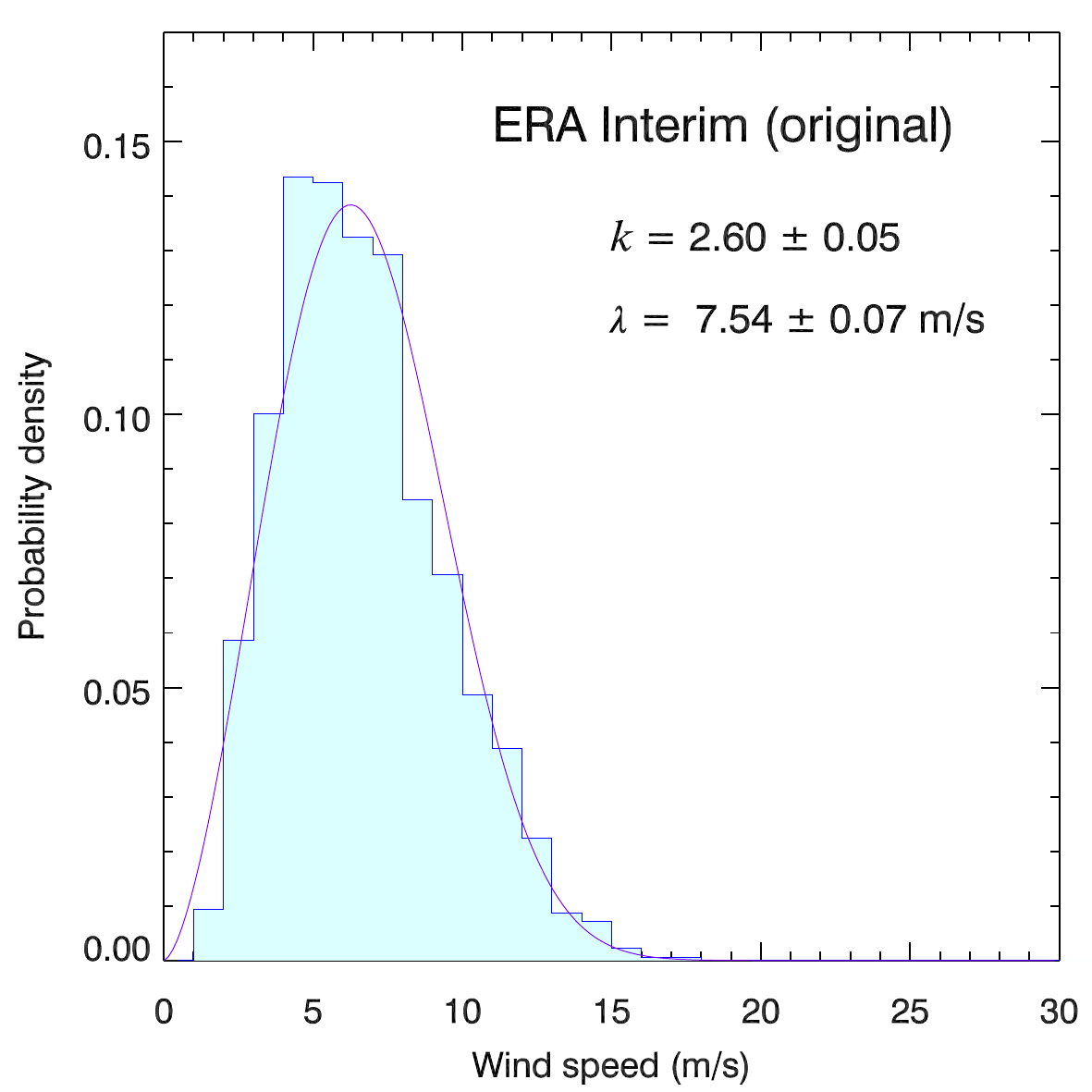}
\includegraphics[width=\figwhere]{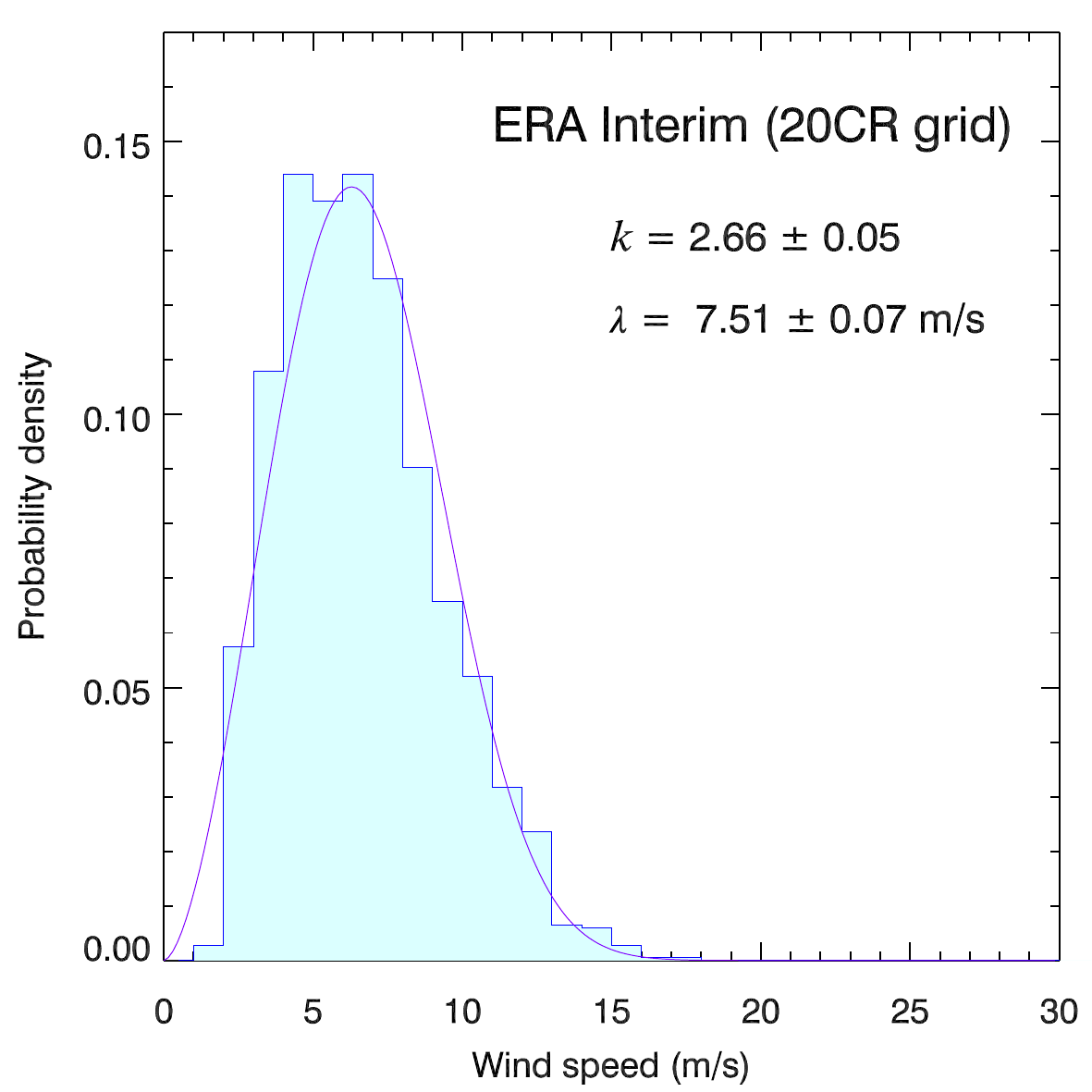} 
\includegraphics[width=\figwhere]{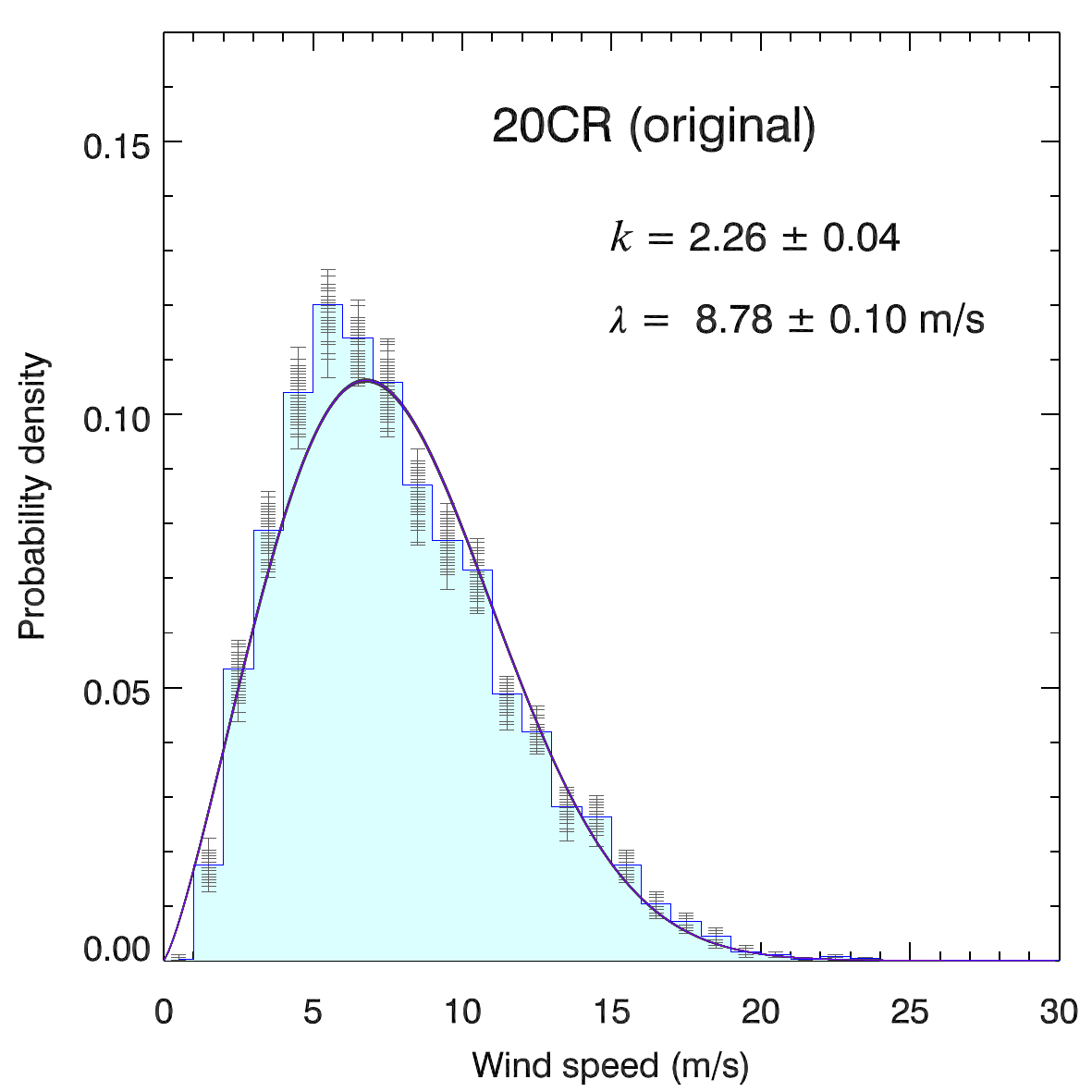}
\includegraphics[width=\figwhere]{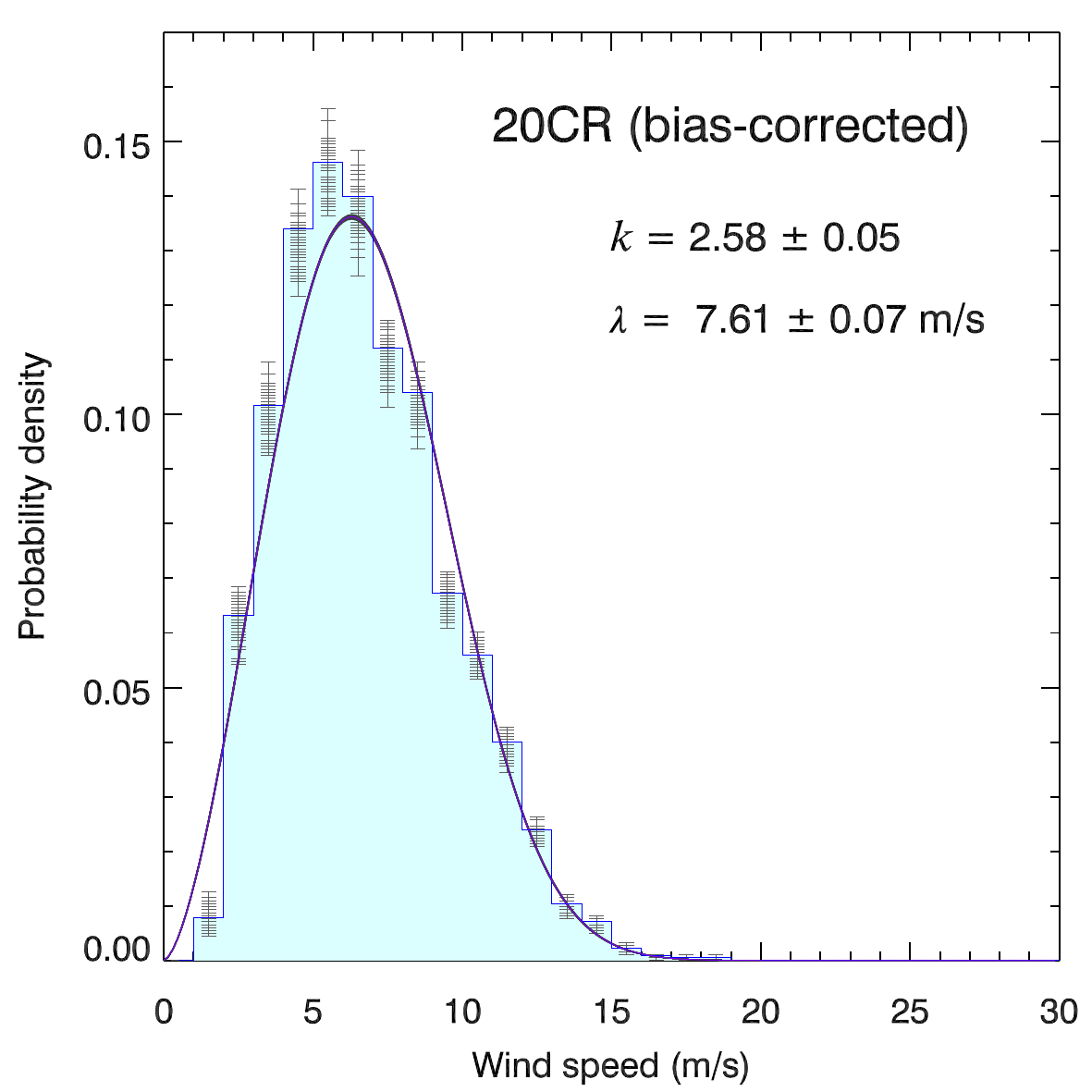}
\caption{Wind speed PDFs for a grid point over western England ($-2\degE$, $52\degN$) for the period of 2001--2005 inclusive. We compare ERA-Interim on its original grid ($0.75\degr$) and on the 20CR grid ($2\degr$) with both the original and calibrated 20CR data.  The ensemble members' histograms are shown as grey horizontal lines, with the ensemble-mean in each bin shown as the blue histogram.   The Weibull fit for each ensemble member is plotted in dark grey, with the Weibull function of the ensemble-mean shape and scale overplotted in purple; these are practically indistinguishable. }
\label{f:weibgofpdfs}
\end{figure*}

\begin{figure} \centering\includegraphics[width=\figw]{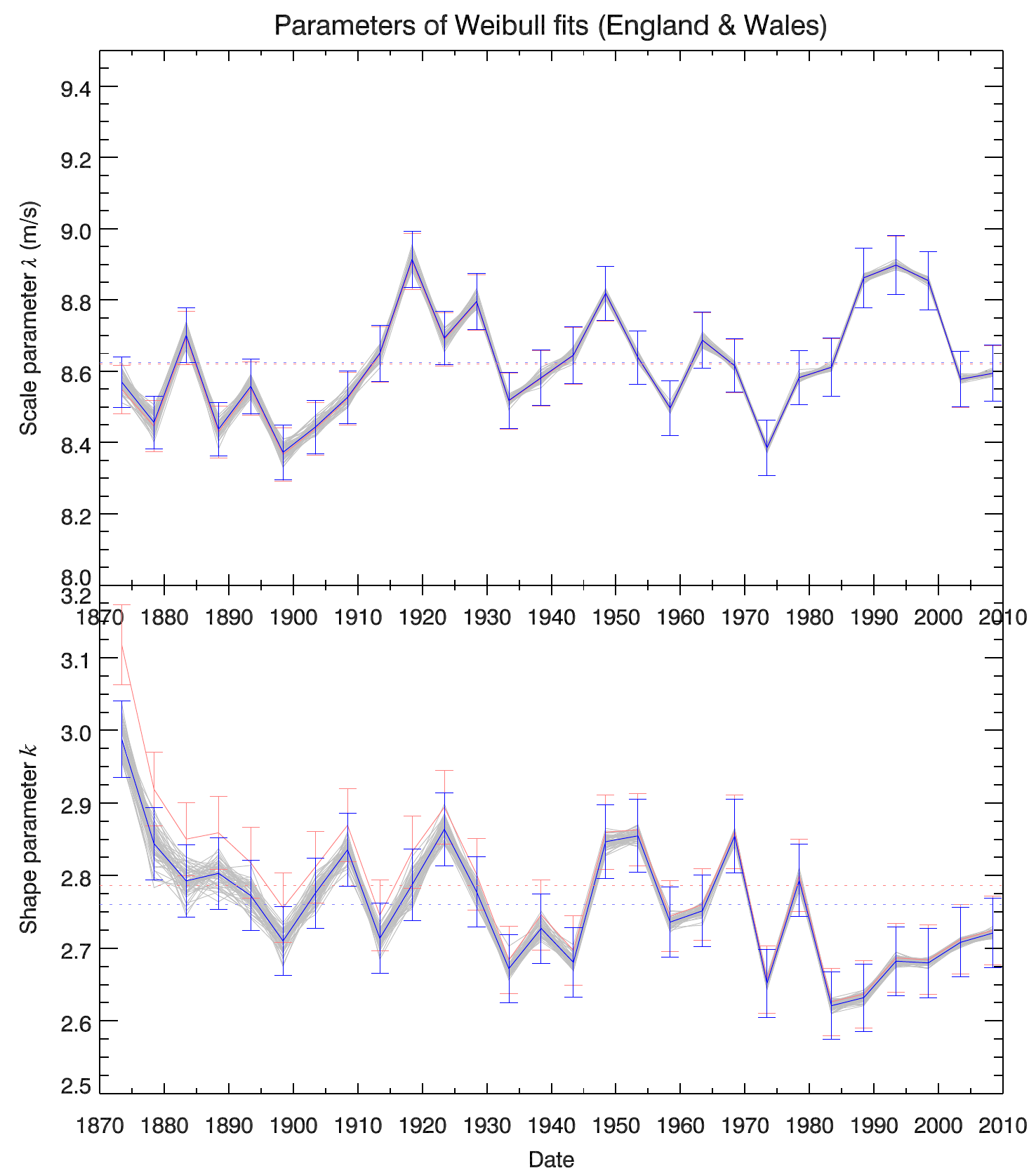}
  \caption{Time series of Weibull shape and scale parameters for the England \& Wales region, in 5-year steps. The results for  individual ensemble members are plotted in grey; despite not being continuous in time, we join their points to aid visibility. The ensemble mean of the parameters is plotted in blue, with error bars showing the ensemble means of the formal 1-$\sigma$ errors on the fits. Plotted in light red are the fits to the daily ensemble-mean time series, and their 1-$\sigma$ errors.  The long-term means of both time series are plotted as dotted lines in blue and red. }
  \label{f:weibts_englwal}
\end{figure}

The time series of Weibull parameters for the England \& Wales region are shown in Figure~\ref{f:weibts_englwal}.  The scale parameter $\lambda$ behaves very similarly to the time series of the mean and standard deviation of wind speeds shown in the main paper (although note that we are looking at discrete 5-year periods here, rather than the 5-year rolling means shown there). The shape parameter $k$ does not vary a great deal, staying between $2.6$ and $3$, although there is noticeable spread at early decades between the values estimated for different ensemble members.    For this region, neither the scale nor shape parameters exhibit trends in their 5-yearly ensemble-mean time series that are significantly different from zero.  Note also that the estimate of the shape parameter from the fits to the daily ensemble mean time series is biased, as expected, compared to the estimates from the fits to the ensemble members individually, and their resulting ensemble mean.

\begin{figure}
  \includegraphics[width=\figw]{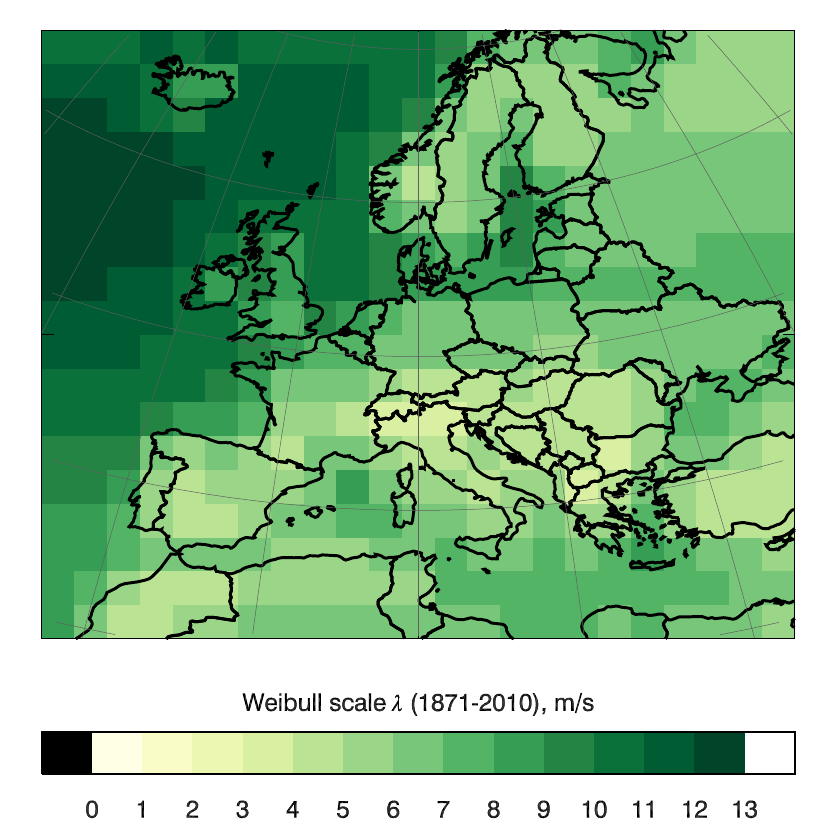}
  \includegraphics[width=\figw]{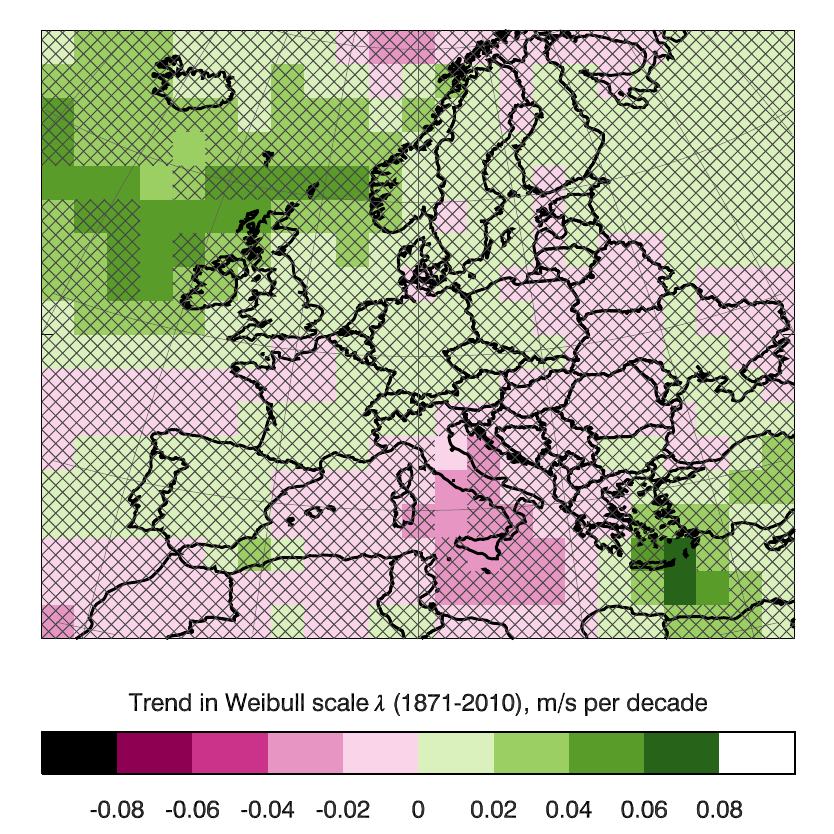}
  \caption{Maps of the Weibull scale parameter.  The top panel shows the long-term value (the time-average of the ensemble means of Weibull fits in 5-year periods), and the bottom panel shows the trend in the series of 5-year values, with crosshatching over cells where the trend is consistent with zero  at the $0.1\%$ level. }
  \label{f:wfitmaps_scale}
\end{figure}

\begin{figure}
  \includegraphics[width=\figw]{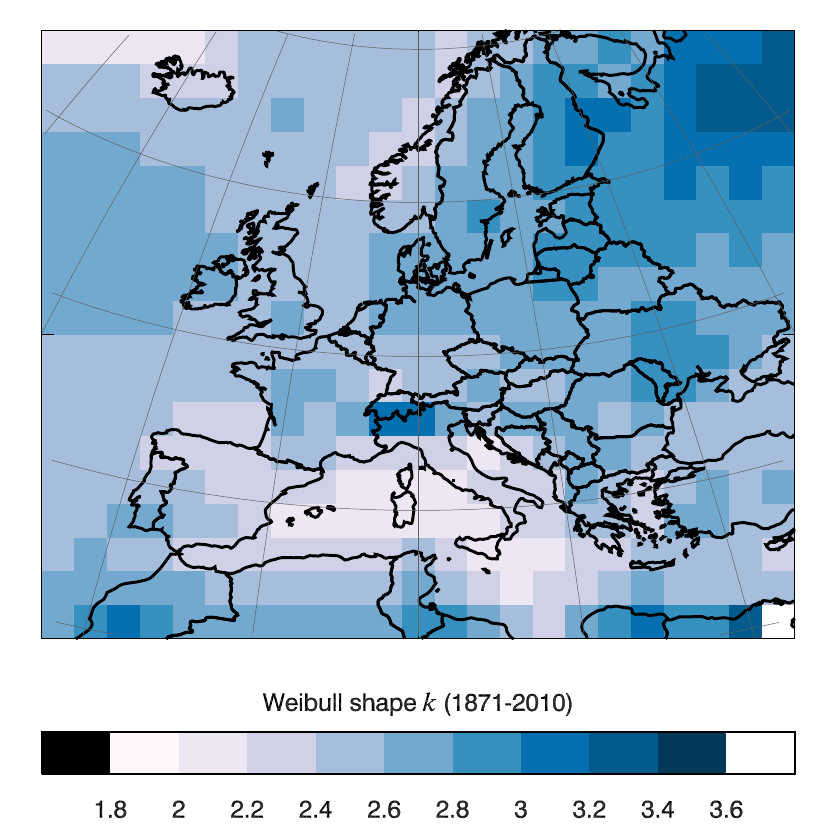}
  \includegraphics[width=\figw]{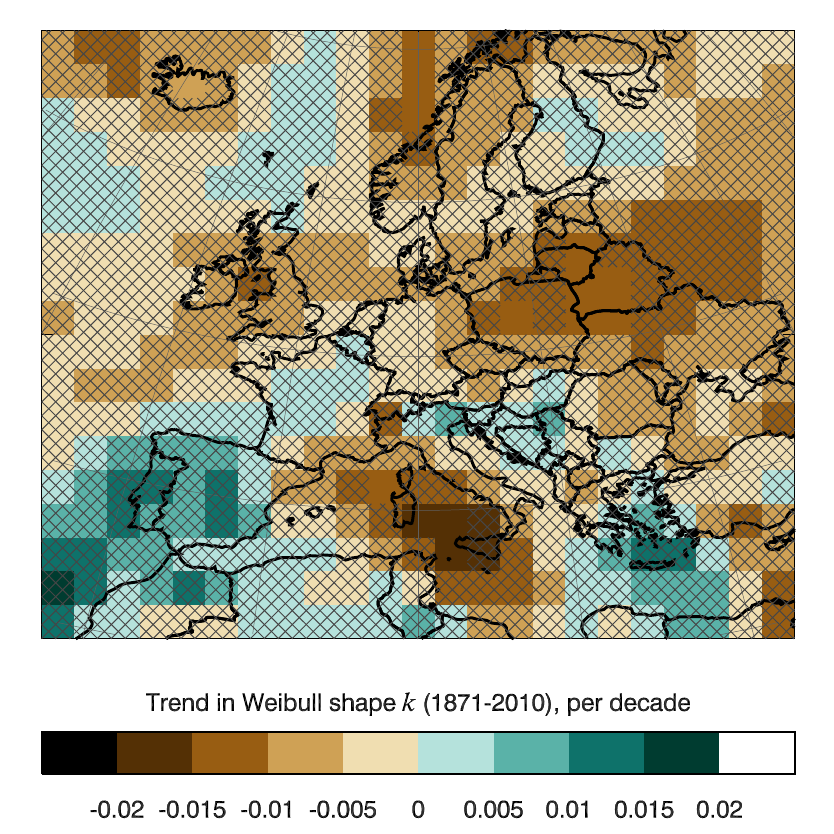}
  \caption{Maps of the Weibull shape parameter.  The top panel shows the long-term value (calculated as in the previous figure), and the bottom panel shows the trend in the series of 5-year values, with crosshatching over cells where the trend is consistent with zero at the $0.1\%$ level. }
  \label{f:wfitmaps_shape}
\end{figure}

Maps of the long-term mean values and trends of the Weibull scale and shape parameters are shown in Figures~\ref{f:wfitmaps_scale} and~\ref{f:wfitmaps_shape}.    The long-term means are calculated as the averages over the 28 5-year periods of the ensemble mean parameter values based on the fits to each member.  The trends are calculated  using the Theil--Sen trend estimator on the 5-yearly ensemble mean values,  and tested for significance using modified Mann--Kendall test at the $0.1\%$ level, as for the mean wind speed time series.

Figure~\ref{f:wfitmaps_scale} shows the mean and trend for the Weibull scale parameter.  As expected, the mean scale map shows a very similar spatial pattern to the long-term mean wind speed map shown in the main paper. The areas with statistically significant trends are  smaller in this case than for the mean wind speed, although the trends themselves have a similar magnitude and spatial pattern. 

The results for the shape parameter are shown in Figure~\ref{f:wfitmaps_shape}. The long-term value of the shape parameter is much less spatially variable than the scale, but in most areas is above the value of $k=2$, i.e. the distribution is less skewed than a Rayleigh distribution.  The only areas with statistically significant trends are in the central Mediterranean -- we have already flagged this region as being subject to data quality issues, and it is discussed further in the Appendix of the main paper. 

\cite{Zhou2013Spatial} studied the Weibull parameters derived from hourly  $10\munit$ winds over land only, using  the NCEP Climate Forecast System Reanalysis \citep[CFSR,][]{Saha2010NCEP} covering 1980--2009.  The values they derive for $\lambda$ and $k$ are noticeably smaller than those we have shown here. This is to be expected, for the scale parameter at least, because of the difference in height compared to our winds at $60\munit$.  However, the CFSR is also at much higher resolution than the 20CR (around $0.3\degr$ compared to $2\degr$), which makes direct comparison difficult.  However, their results are similar to those of  \cite{Kiss2008}, who used the 6-hourly $10\munit$ winds from the ERA-40  reanalysis \citep{Uppala2005}, covering 1958--2002 at a resolution of $1\degr$.  \cite{Kiss2008}, like us, also noted that Weibull fits tend to be much better over the ocean than the land; the data used by \cite{Zhou2013Spatial} only covers land area.  \cite{BettThorntonClark2013}, considered the uncorrected 20CR data using magnitudes of daily mean wind vectors (unlike our present study). In that case, slightly lower values for Weibull scale were obtained compared to here (as expected from our calibration usually acting to increase wind speeds), and slightly smaller Weibull shapes (i.e. less symmetric distributions before calibration, as expected from Figure~\ref{f:weibgofpdfs}).  In that paper, the trend analysis was performed using simple linear regression and a $t$-test at $0.1\%$ significance; the magnitudes and spatial patterns of the trends are very similar to our present results, although we find smaller areas of significance.

Finally, it is important to note that being well-described by a Weibull distribution should not be taken as a necessary  indicator of physical realism. The work of \cite{Harris2014Parent} suggests that in regions with strongly varying wind speeds (such as in different seasons), or between regions with different behaviour (such as coastal vs. inland), then a combination of Weibull distributions for the different climatological scenarios (regions/seasons) would be a better fit -- although one then starts to lose the benefit of simplicity and conciseness that led to the Weibull distribution in the first place.

%============================================================================

\end{document}